\documentclass[prb,showpacs,superscriptaddress,preprintnumbers,amssymb,twocolumn,amsfonts]{revtex4-1}
\usepackage{graphicx}
\usepackage{amssymb,amsmath, amsfonts, bm, braket}
\usepackage{dsfont}
\usepackage{subfigure}
\usepackage{color}
\usepackage[breaklinks]{hyperref}

\begin{document}

\def\beq{\begin{equation}}
\def\eeq{\end{equation}}
\def\beqn{\begin{eqnarray}}
\def\eeqn{\end{eqnarray}}
\def\etal{\emph{et al.}}
\def\ket#1{\vert #1 \rangle}
\def\bra#1{\langle #1 \vert}
\def\ev#1{\langle #1 \rangle}
\def\ip#1#2{\langle #1 \vert #2 \rangle}
\def\me#1#2#3{\langle #1 \vert #2 \vert #3 \rangle}
\renewcommand{\bf}{\mathbf}
\newcommand{\Tr}{\operatorname{Tr}}

\definecolor{purple}{rgb}{0.5,0,0.5}
\definecolor{dkgreen}{rgb}{0,0.5,0}
\newcommand{\jonas}[1]{ { \color{dkgreen} \footnotesize (\textsf{JK}) \textsf{\textsl{#1}} }}
\newcommand{\mike}[1]{ { \color{blue} \footnotesize (\textsf{MZ}) \textsf{\textsl{#1}} }}
\newcommand{\roger}[1]{ { \color{purple} \footnotesize (\textsf{RM}) \textsf{\textsl{#1}} } }
\newcommand{\jens}[1]{ { \color{cyan} \footnotesize (\textsf{JHB}) \textsf{\textsl{#1}} }}
\newcommand{\frank}[1]{ { \color{red} \footnotesize (\textsf{FP}) \textsf{\textsl{#1}} }}

\title{The phase diagram of the anisotropic spin-2 XXZ model: an infinite system DMRG study}

\author{Jonas~A.~Kj\"{a}ll}
\affiliation{Department of Physics, University of California, Berkeley, California 94720, USA}
\affiliation{Max-Planck-Institut f\"{u}r Physik komplexer Systeme, 01187 Dresden, Germany}
\author{Michael P. Zaletel}
\affiliation{Department of Physics, University of California, Berkeley, California 94720, USA}
\author{Roger~S.~K.~Mong}
\affiliation{Department of Physics, University of California, Berkeley, California 94720, USA}
\affiliation{Department of Physics, California Institute of Technology, Pasadena, CA 91125, USA}
\author{Jens~H.~Bardarson}
\affiliation{Department of Physics, University of California, Berkeley, California 94720, USA}
\affiliation{Materials Sciences Division, Lawrence Berkeley National Laboratory, Berkeley, California 94720, USA}
\author{Frank~Pollmann}
\affiliation{Max-Planck-Institut f\"{u}r Physik komplexer Systeme, 01187 Dresden, Germany}
\date{\today}

\begin{abstract}
We study the ground state phase diagram of the quantum spin-2 XXZ chain in the presence of on-site anisotropy using a matrix-product state based infinite system density-matrix-renormalization-group (iDMRG) algorithm. One of the interests in this system is in connecting the highly quantum mechanical spin-1 phase diagram with the classical $S=\infty$ phase diagram. Several of the recent advances within DMRG make it possible to perform a detailed analysis of the whole phase diagram. We consider different types of on-site anisotropies which allows us to establish the validity of the following statements: 
One, the spin-2 model can be tuned into a phase which is equivalent to the ``topologically nontrivial'' spin-1 Haldane phase. Two, the spin-2 Haldane phase at the isotropic Heisenberg point is adiabatically connected to the ``trivial'' large-$D$ phase, with a continuous change of the Hamiltonian parameters. Furthermore, we study the spin-3 XXZ chain to help explain the development of the classical phase diagram. We present details on how to use the iDMRG method to map out the phase diagram and include an extensive discussion of the numerical methods.
\end{abstract}
\maketitle

\section{Introduction}
Quantum spin chains have proven to be extremely useful model systems for the study of strongly correlated quantum systems. In particular, many different types of phases and phase transitions can be understood by studying relatively simple Hamiltonians. A prime example is the SU(2) symmetric Heisenberg chain; it has gapless excitations for half integer spins, while the ground state of integer spin chains is protected by a gap in the energy spectrum. The existence of this gap was predicted by Haldane in the early eighties in Refs.~\onlinecite{HaldaneA,HaldaneB}.  Following Haldane's predictions, Affleck, Kennedy, Lieb, and Tasaki (AKLT) presented model Hamiltonians for which the ground state can be obtained exactly.~\cite{AKLT,AKLT-long} The AKLT state was later found to exhibit unexpected properties, such as a nonlocal ``string order'' and edge states, which extend to other states within the same phase.~\cite{Nijs,Kennedy-1990,Kennedy-1992,Oshikawa}
The phase including the SU(2) symmetric point has consequently been referred to as the ``Haldane phase''. 

These gapped ground states do not break any symmetry and thus cannot be characterized by any local order parameter. Instead, they are characterized by the projective representations of the symmetries present.\cite{Berg-2008,Gu-2009,Pollmann-2010,Fidkowski-2010,Turner-2010,Pollmann12,ChenGu-2011,ChenGu-2011-2,Schuch-2011} Physically, this means that the spin fractionalizes into two \emph{half-integer} edge spins in the case of odd integer spin chains (one on each edge), and into two \emph{integer} edge spins in the even integer spin chains. 
There is a crucial difference between these two cases.\cite{Pollmann12} In the odd integer case, the Haldane phase is a \emph{symmetry-protected topological phase} (SPTP) as the half-integer edge spin cannot be removed unless the system undergoes a phase transition or all the relevant symmetries are explicitly broken. In contrast, in the even case the integer spins at the edge are not protected and thus the ground state can be adiabatically turned into a trivial (product) state, without breaking any symmetries.
This motivates the notation of two distinct phases, an odd-Haldane (OH) and an even-Haldane (EH) phase, which we adopt here. Seemingly, this would suggest the absence of a topological phase in the even integer spin chains. However, as proposed by Oshikawa,\cite{Oshikawa} all the Haldane phases corresponding to lower integer spin can in principle be realized in the presence of on-site anisotropy. In the spin-2 case, in particular, this implies the presence of the OH phase. 
It is one of the main goals of this work to verify if, and under what conditions, this scenario is realized.

The study of such questions has been greatly advanced by the development of matrix-product-states (MPS)~\cite{Fannes-1992,OstlundRommer1995,RommerOstlund1997} and the reinterpretation of density-matrix-renormalization-group (DMRG) algorithms in terms of these states.~\cite{White-1992,Schollwoeck:Review95,Schollwoeck11,McCulloch-2007} Infinite-system algorithms directly obtain the thermodynamic limit of infinite system size, without any finite-size corrections. This is especially crucial in the vicinity of critical phases, as will turn out to be the case in this work, with diverging correlation lengths. We adopt the infinite-system DMRG, or iDMRG, algorithm in this paper.~\cite{McCulloh-2008,Crosswhite-2008} Although this algorithm has been discussed priorly in the literature, we believe it is beneficial to give a detailed, pedagogical account of the algorithm and to compare and contrast it to the related infinite-system time-evolving-block-decimation (iTEBD) algorithm.~\cite{Vidal-2003a,Vidal-2004,Vidal-2007} This constitutes another main goal of this work.

The remainder of the paper is organized as follows: We begin by presenting the model and reviewing relevant prior work in Sec.~\ref{sec:model}. We present our spin-2 phase diagram and summarize its most important aspects in a summary of our main results in Sec.~\ref{sec:resultsummary}. In Sec.~\ref{sec:MPS} we formulate the iTEBD and iDMRG algorithms using a consistent notation. This allows us to highlight the similarities and differences of the algorithms. In Sec.~\ref{sec:QPT} we review the phases and phase transitions present in the model and discuss how they are most accurately observed with iDMRG. Details of our results, including data for an SPTP in a spin-2 chain and the adiabatic connection of the AFM Heisenberg point and points at large-$D$, are presented in the same section. We wrap up our investigation by presenting the first, to our knowledge, numerical results for the spin-3 XXZ-chain, in Sec.~\ref{sec:S3}. 

\section{Model}
\label{sec:model}

\subsection{Definitions and prior studies}
\begin{figure}[tb!]
  \begin{center}
    \subfigure{\includegraphics[width=70mm]{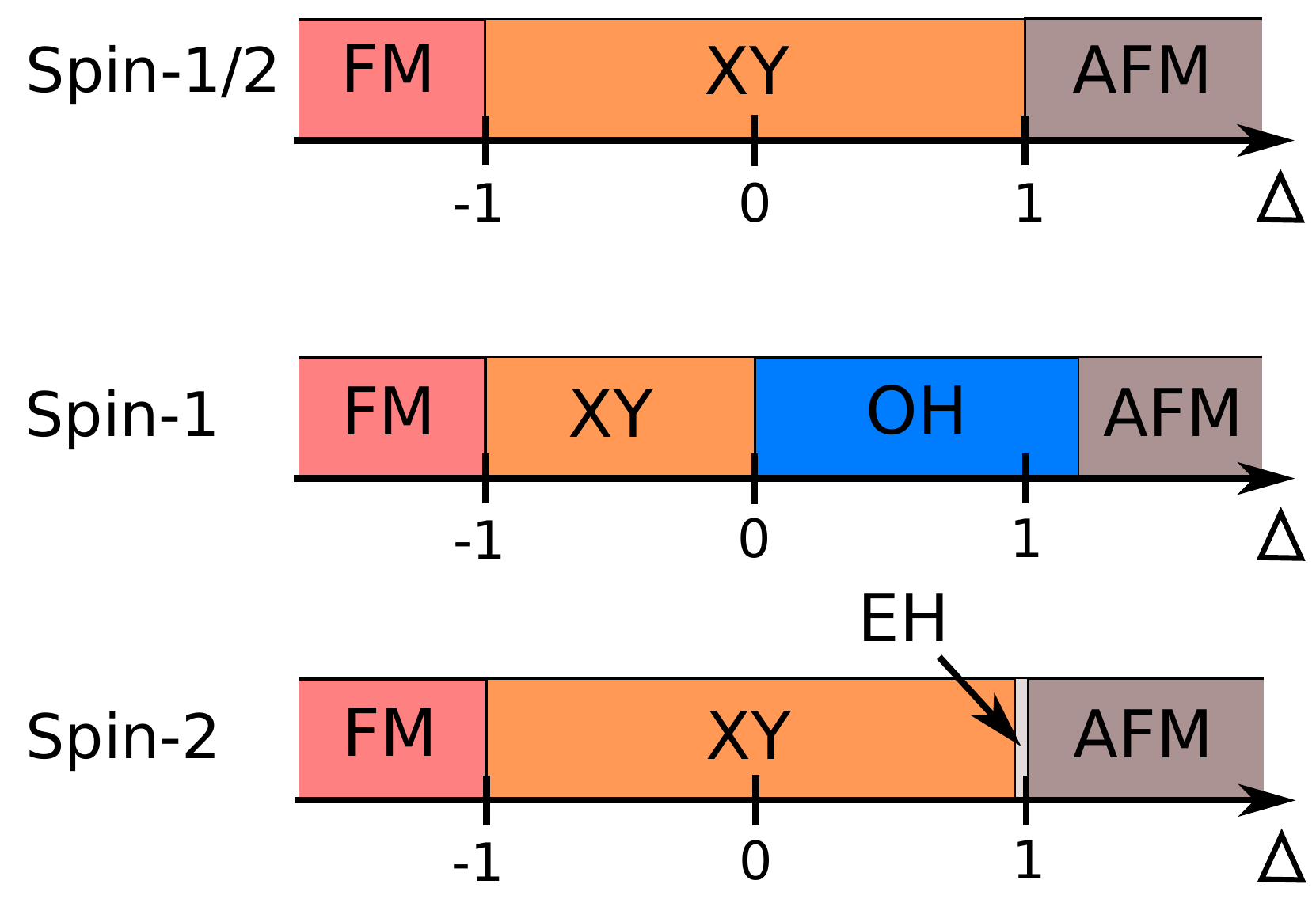}}\\
    \caption{(Color online) The phases present in a spin-$S$ XXZ-chain, for $S=\frac{1}{2},1,2$. The Haldane phases appear only for integer spins and their width decreases rapidly with increasing $S$. At the same time, the phase transitions into the Haldane phases approach $\Delta=1$, where the direct XY-AFM phase transition occurs for all half-integer $S$.}
    \label{XXZpd}
  \end{center}
\end{figure}

The model Hamiltonian we are concerned with in this work is of the form
\begin{equation}
	H = H_{\rm XXZ} + H_D.
	\label{Hamiltonian}
\end{equation}
The first term describes the XXZ quantum spin chain 
\begin{equation}
H_{\text{XXZ}}=J\displaystyle\sum_n \left (S^x_nS^x_{n+1}+S^y_nS^y_{n+1}+\Delta S^z_nS^z_{n+1}\right),
\label{XXZ}
\end{equation}
where $S^{\alpha}_n$, with $\alpha =x,y,z$, is the $\alpha$-component of the spin-$S$ operator at site $n$. $\Delta$ is the XXZ anisotropic interaction parameter. As $\Delta$ is tuned from $-\infty$ to $\infty$,
the following phases appear: ferromagnetic (FM), XY, Haldane (OH/EH, for integer spins only) and anti-ferromagnetic (AFM), see Fig.~\ref{XXZpd} for $S=\frac{1}{2},1,2$.~\cite{HaldaneA,HaldaneB,Giamarchibook} In the following we will set the overall energy scale to $J=1$. The transition into the FM phase occurs at $\Delta=-1$ for all $S$. The Heisenberg point $\Delta=1$ is the transition into the AFM phase for half integer $S$. For integer $S$ it is within the Haldane phase, whose width in $\Delta$ decreases rapidly with increasing $S$ consistent with its absence in the classical $S\rightarrow \infty$ phase diagram.~\cite{HaldaneA,HaldaneB} This makes it computationally demanding to resolve the Haldane phases for large S.

The term $H_D$ represents an on-site anisotropy, which for spin $S$ has the general form
\begin{equation}
H_{\text{D}}=\sum_n\sum_p^{2S} D_p(S^z_n)^p.
\label{Onsite}
\end{equation} 
with $D_p$ constants. Depending on the values of the $D_p$, this on-site anisotropy can favor all the different eigenstates $|m_{\alpha}\rangle$ of a noninteracting spin $\vec{S}$, and allows for realization of more phases than the XXZ model alone. The terms with odd power tend to favor the magnetically ordered AFM and FM phases, so we do not include them here. In the case of spin-2, there are then two terms with coupling constants $D_2$ and $D_4$. A large positive $D_2$ (or a large $D_4$) favors a product state of the form
\begin{equation}
	|\psi_D\rangle = \bigotimes_n |0\rangle_n,
	\label{eq:productstate}
\end{equation}
where $|0\rangle_n$ is the eigenstate of $S^z_n$ with eigenvalue $m_z = 0$. This is because  the higher $m_z$ eigenstates have been energetically projected out, and there remains only an effective spin-0 degree of freedom. In the limit $D_4\rightarrow\infty$ with $D_2 = -D_4$,  only the states with $m_z = \pm 2$  are projected out, and an effective spin-1 degree of freedom remains. This allows for the exploration of spin-1 phases, such as the OH phase, in the spin-2 chain. While nonzero $D_2$ has been considered before, see Ref.~\onlinecite{Schollwoeck95,Schollwoeck96,Aschauer,Nomura,Tonegawa,Tzeng}
, we are not aware of any study that also includes the higher order $D_4$ term.

\begin{figure}[tb!]
  \begin{center}
    \includegraphics[width=70mm]{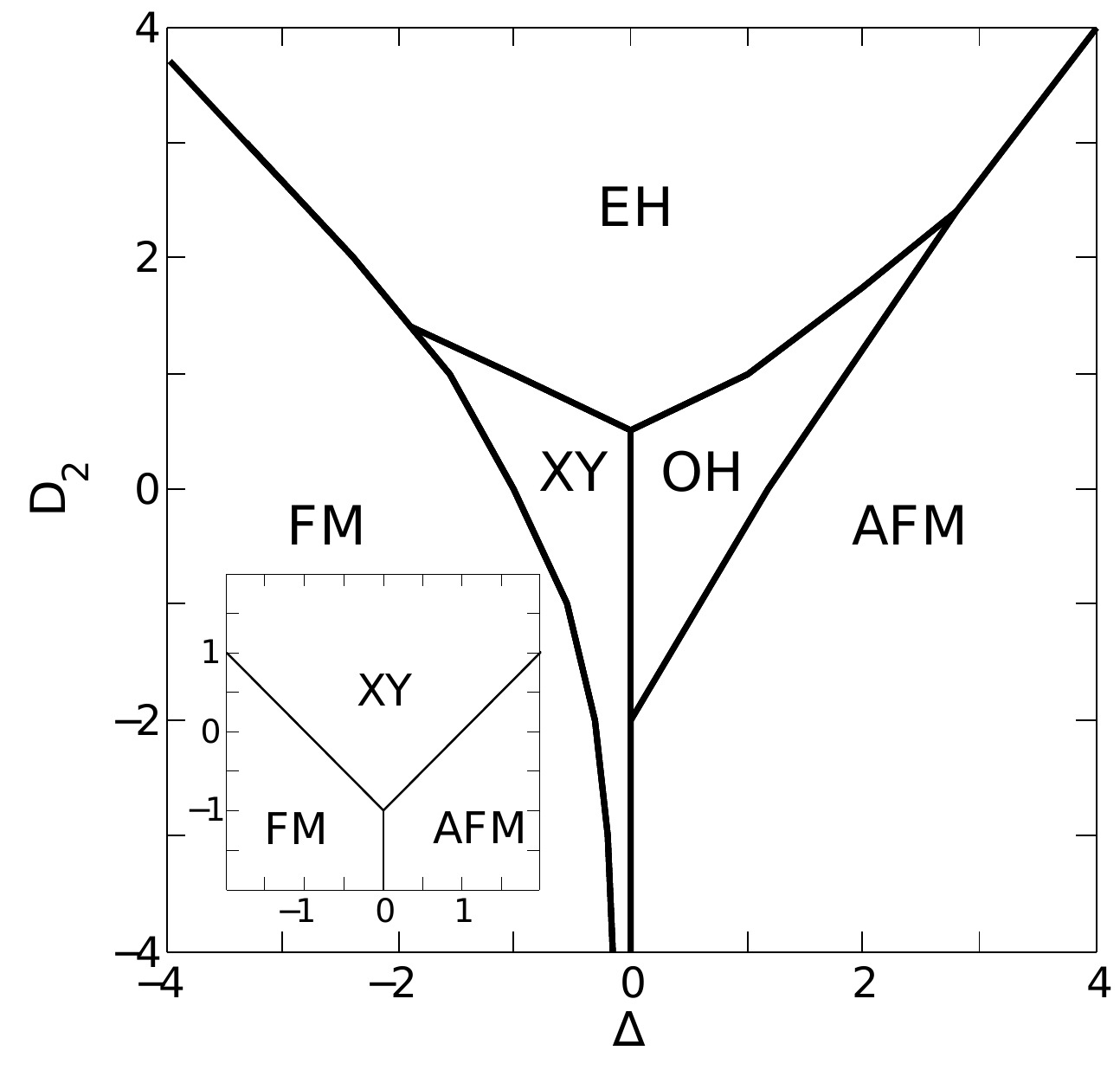}
    \caption{A sketch of the spin-1 phase diagram from Ref.~\onlinecite{ChenLS} (with our labeling of the phases). Inset: Phase diagram in the semiclassical limit $S\rightarrow\infty$.}
    \label{S1pd}
  \end{center}
\end{figure}

We start by revisiting the case with $D_2$ as the only nonzero on-site anisotropy.  For spin-1, the phase diagram is well established\cite{Botet,Schulz,SchulzZiman,Nijs,Kitazawa96,ChenLS}
and is schematically plotted in Fig.~\ref{S1pd}. With increasing $D_2$, there is a transition into a phase that contains the trivial product state~\eqref{eq:productstate} as a ground state. This phase is often called the `large-$D$' phase. However, in light of the recent classification scheme in terms of SPTP\cite{Berg-2008,Gu-2009,Pollmann-2010,Fidkowski-2010,Turner-2010,Pollmann12,ChenGu-2011,ChenGu-2011-2,Schuch-2011}, the large-$D$ phase is expected to be representative of the EH phase, so we will denote it as such (in fact, it can be thought of as the trivial $S=0$ Haldane phase).

	The situation for $S\geq 2$ is not as clear. Bosonization predicts that the spin-1 phase diagram is representative of all integer spins\cite{Schulz} for small $D_2$. Oshikawa predicted intermediate phases, corresponding to all the phases at the Heisenberg point for lower integer spins, between the Haldane phase and the large-$D$ phase.~\cite{Oshikawa} Early DMRG studies\cite{Schollwoeck95,Schollwoeck96,Aschauer} did not corroborate this picture but found for $S=2$ a phase diagram already close to the semi-classical limit $S\rightarrow\infty$ (inset Fig.~\ref{S1pd}). The Haldane phase surrounding the Heisenberg point appeared to be \emph{separated} from the trivial large-$D$ phase by the XY phase. 
Later level spectroscopy (LS) studies\cite{Tonegawa,Tzeng} suggested instead that the Haldane phase is connected with the large-$D$ phase; furthermore, these studies concluded in favor of the presence of a tiny OH phase in the $D_4 = 0$ plane around ($\Delta=2.4,D_2=2$). 

The two main questions, that have  been hard to verify for the spin-2 case, and which we address here are: i) Is the Heisenberg point indeed adiabatically connected to the trivial large-$D$ phase? ii) Is there an OH phase in the phase diagram, particularly in the experimentally relevant case of small on-site anisotropy? To answer these questions, we will find it useful to introduce nonzero $D_4$, as this will allow us to easily access the OH phase and determine its extent.

\section{Summary of the main results} 
\label{sec:resultsummary}
\begin{figure}[tb!]
  \begin{center}
    \subfigure{\includegraphics[width=70mm]{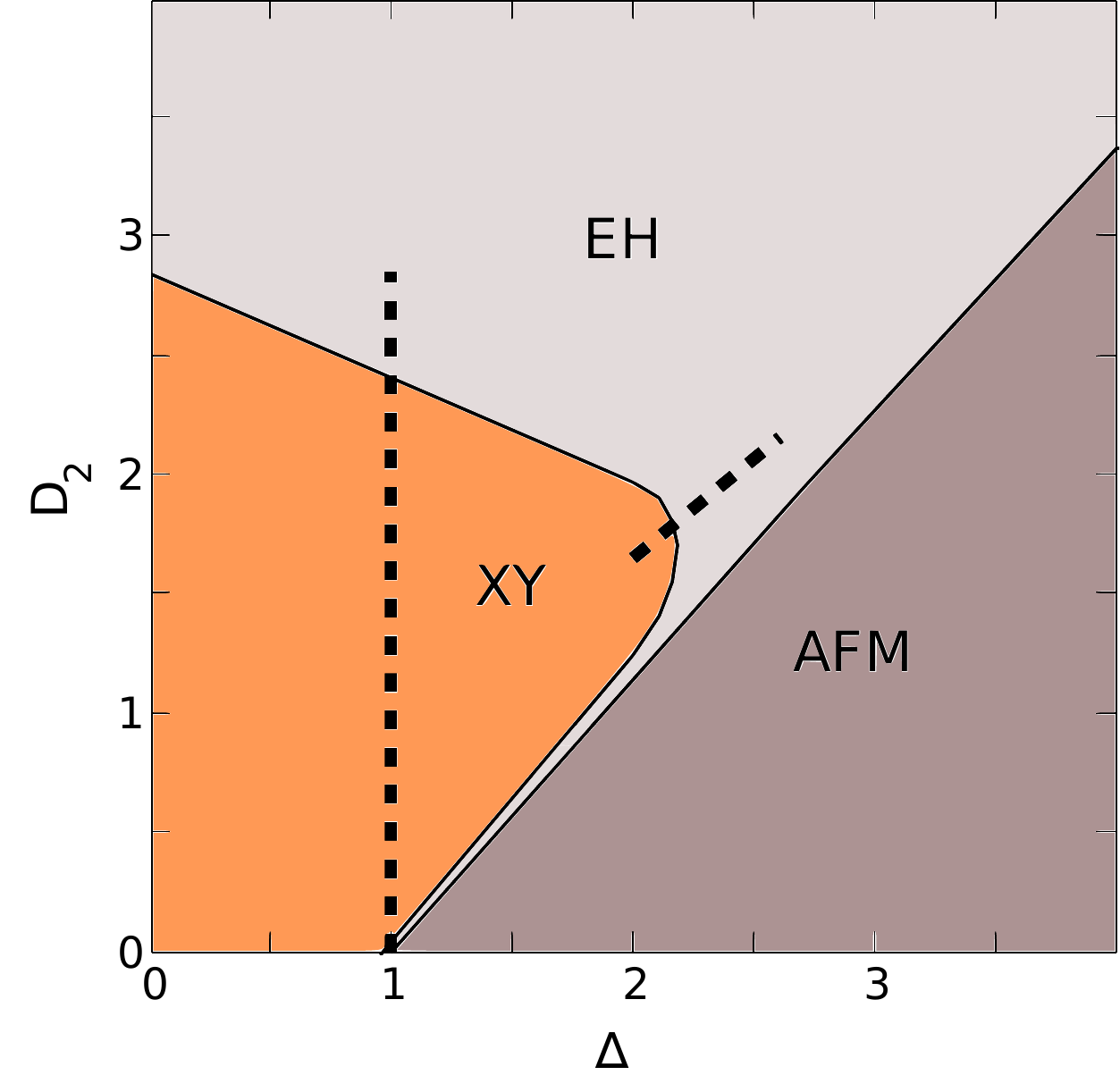}}\\
    \caption{(Color online) The spin-2 phase diagram at $D_4=0$ for $\Delta\geq 0$ and $D_2\geq 0$ as obtained with iDMRG. Examples of data used to obtain the phase boundaries are presented in Sec.~\ref{sec:QPT}. In particular, data corresponding to the vertical line is given in Figs.~\ref{AFMEH} and~\ref{BKT} and the tilted line in Figs.~\ref{BKT} and~\ref{Gaus}.}
    \label{JzD2pd}
  \end{center}
\end{figure}

\begin{figure}[tb!]
  \begin{center}
    \subfigure{\includegraphics[width=75mm]{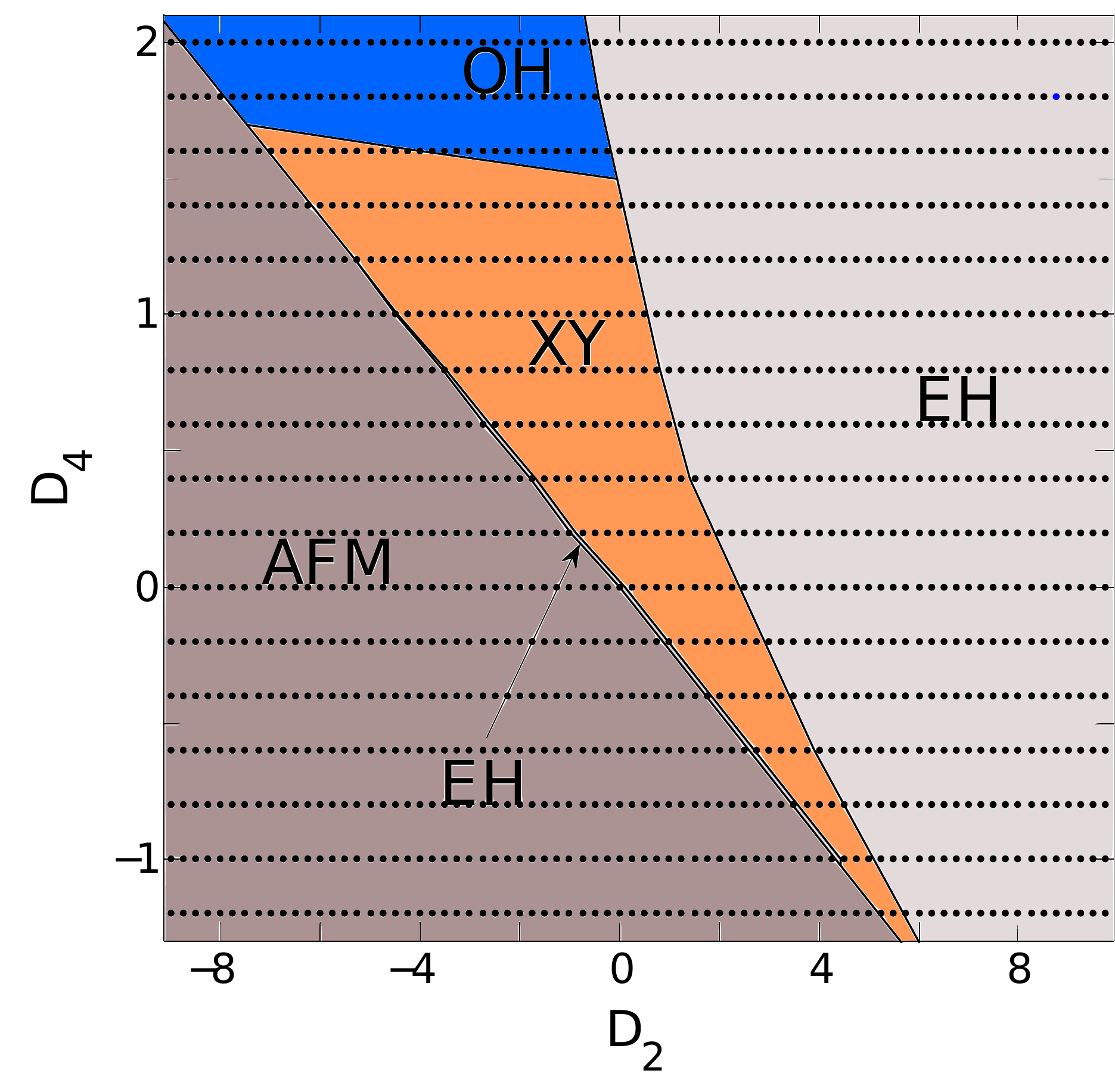}}\\
    \caption{(Color online) The spin-2 phase diagram at $\Delta=1.0$ as obtained with iDMRG. The OH phase is easily reached by including the $D_4$ on-site anisotropy term. The dots indicate the values at which the simulations were performed. To resolve the narrow EH phase, simulations on a finer grid were performed between the AFM and XY phases (points not shown).
    }
    \label{D2D4pd}
  \end{center}
\end{figure}

In this section we summarize our main $S=2$ results by discussing the phase diagrams obtained in our iDRMG simulations and shown in Fig.~\ref{JzD2pd} and Fig.~\ref{D2D4pd}, deferring the details of how they were obtained to Sec.~\ref{sec:QPT}.
The phase diagram for the relevant upper right corner $\Delta\geq 0,D_2\geq 0$ at $D_4=0$ is shown in~Fig.~\ref{JzD2pd}.  
There is no direct transition from the XY phase into the AFM phase.
Instead, these phases are separated everywhere by the EH phase which is continuously present to large $D_2$. This answers the first of our main questions; the Heisenberg point and the large-$D$ phase are continuously connected.

The extension of the XY phase is consistent with the recent LS study of Ref~\onlinecite{Tonegawa}, but covers a much smaller range of $\Delta$ than obtained in the earlier DMRG study of Ref.~\onlinecite{Aschauer}. However, our numerical results suggest the complete absence of an OH phase in this plane, in contrast to Refs.~\onlinecite{Tonegawa, Tzeng} where it is observed next to the XY phase. We believe that the reason for these different conclusions arises from the proximity of an EH $\leftrightarrow$ OH phase transition at small $D_4$, as discussed below, which makes the relevant parameter region appear critical in finite chains.

The $D_4$ anisotropy is an important parameter in establishing the above result. In fact, since large $D_4 = -D_2$  effectively realizes a spin-1 chain, one expects the OH phase to appear at the Heisenberg point along the lines $D_2=-D_4+D_2^{S=1}$ with $D_2^{S=1}$ the values at which the OH phase appears in the $S=1$ phase diagram. This is indeed what we observe, as shown in Fig.~\ref{D2D4pd}. The fact that the OH phase is easily realized when introducing the additional anisotropy parameter $D_4$ is one of the main new result of our work. The question of whether it is obtained in the $D_4 = 0$ plane is then simply the question of how large the extent of the OH phase is. While we believe that our results demonstrate that it does not touch the $D_4 = 0$ plane, we cannot strictly rule out that it does, as discussed in more detail in Sec.~\ref{sec:QPT}. 

\section{Numerical Techniques}
\label{sec:MPS}

In this section we outline the numerical method used to obtain our results. Though most of the details have been discussed in the literature elsewhere, we find it useful to discuss our implementation of the algorithm. The infinite time evolving block decimation (iTEBD)~\cite{Vidal-2007} and the infinite density matrix renormalization group (iDMRG)~\cite{McCulloch-2007} algorithms are both based on the infinite matrix-product state (iMPS) representation.~\cite{Vidal-2007} As we explain shortly, MPSs can efficiently represent many-body wave functions where the accuracy is controlled by the \emph{bond dimension} $\chi$ (the error decreases rapidly with increasing $\chi$).
Using methods which work with infinite systems has a number of advantages:  no extrapolation to the thermodynamic limit is needed; there are no edge modes which can complicate the convergence of the algorithm; and, as shown later on in this section, finite entanglement scaling can be used to extract quantities such as the central charge. We begin by reviewing some details of this infinite system representation focusing on translationally invariant systems, and then contrast and compare the two numerical methods using a consistent notation. We do not aim to provide a complete discussion of the techniques but rather a clear and compact introduction to the methods used.

The concept of entanglement is central to the MPS representation and the algorithms based on it. The so-called  \emph{entanglement spectrum} \cite{Li-2008} is obtained from the Schmidt decomposition (singular value decomposition): Given a bipartition $\mathcal{H} = \mathcal{H}_L \otimes \mathcal{H}_R$ of the Hilbert space (below $\mathcal{H}_L$ and $\mathcal{H}_R$, respectively, represent the states on sites to the left and right of a bond),  any state $\ket{\Psi}\in\mathcal{H}$ can be decomposed as
\begin{align}
	\ket{\Psi} &= \sum_{\alpha} \Lambda_\alpha \ket{\alpha}_L \otimes \ket{\alpha}_R, \quad
		\ket{\alpha}_{R/L} \in \mathcal{H}_{R/L} .
\end{align}
The Schmidt coefficients (singular values) $\Lambda_\alpha$ can always be chosen positive, the states $\{\ket{\alpha}_L\}$ and $\{\ket{\alpha}_R\}$ form orthonormal sets in $\mathcal{H}_L$ and $\mathcal{H}_R$ respectively, i.e., $\braket{\alpha|\beta}_L = \braket{\alpha|\beta}_R= \delta_{\alpha\beta}$, and by normalization $\sum_\alpha \Lambda_\alpha^2 = \braket{\Psi|\Psi} = 1$. The Schmidt decomposition is related to the reduced density matrix for one half of the system, $\rho^{R} = \mathrm{Tr}_{\mathcal{H}_L} \left( \ket{\psi}\bra{\psi} \right)$. In particular, the Schmidt states are the eigenstates of $\rho^R$ and the Schmidt coefficients are the square roots of the corresponding eigenvalues, i.e.,  
$\rho^{R} = \sum_\alpha \Lambda^2_{\alpha} \ket{\alpha} \bra{\alpha}_R$ (and analogously for $\rho^L$). This directly gives the entanglement entropy through 
\begin{equation}
S_E = -\sum_\alpha \Lambda_\alpha^2 \log \Lambda_\alpha^2.
\label{eq:SE}
\end{equation}
Finally, the entanglement spectrum $\{\epsilon_\alpha\}$ is related to the spectrum $\{\Lambda_{\alpha}^2\}$ of the bipartition by $\Lambda_\alpha^2 = \exp(-\epsilon_\alpha)$ for each $\alpha$.

\subsection{Matrix-product states}
A general quantum state $|\Psi\rangle$ on a chain with $N$ sites can  be written in the following MPS form:
\cite{Fannes-1992,OstlundRommer1995,RommerOstlund1997}
\begin{equation}
	|\Psi \rangle = \sum_{j_1, \ldots, j_N} A^{[1]j_1}A^{[2]j_2} \ldots A^{[N]j_N} | j_1, \ldots ,j_{N} \rangle.  \label{eq:mps}
\end{equation}
Here, $A^{[n]j_n}$ is a $\chi_{n-1} \times \chi_{n}$ matrix and $|j_n\rangle$ with $j_n=1,\dots,d$ is a basis of local states at site $n$.
We call the indices of the matrices ``bond'' indices.
The matrices at the boundary, i.e., $n=1$ and $n=N$, are vectors, that is $\chi_0 = \chi_{N} = 1$, such that the matrix product in~\eqref{eq:mps} produces a scalar coefficient.
The superscript $[n]$ denotes the fact that for a generic state, each site is represented by a different set of matrices. 
Ground states of one dimensional gapped systems can be efficiently approximated by an MPS,\cite{Gottesman-2009, Schuch-2008} in the sense that the value of the $\chi$'s needed to approximate the ground state wave function to an arbitrary precision is finite as $N\rightarrow \infty$. The physical insight that allows us to make this statement is the area law, which holds for this class of systems.~\cite{Verstraete-2006,Hastings-2007} Details on how the accuracy of the representations depends on $\chi$ can be found in Ref.~\onlinecite{Verstraete-2006}. 


\begin{figure}[tb!]
	\includegraphics[width=8.5cm]{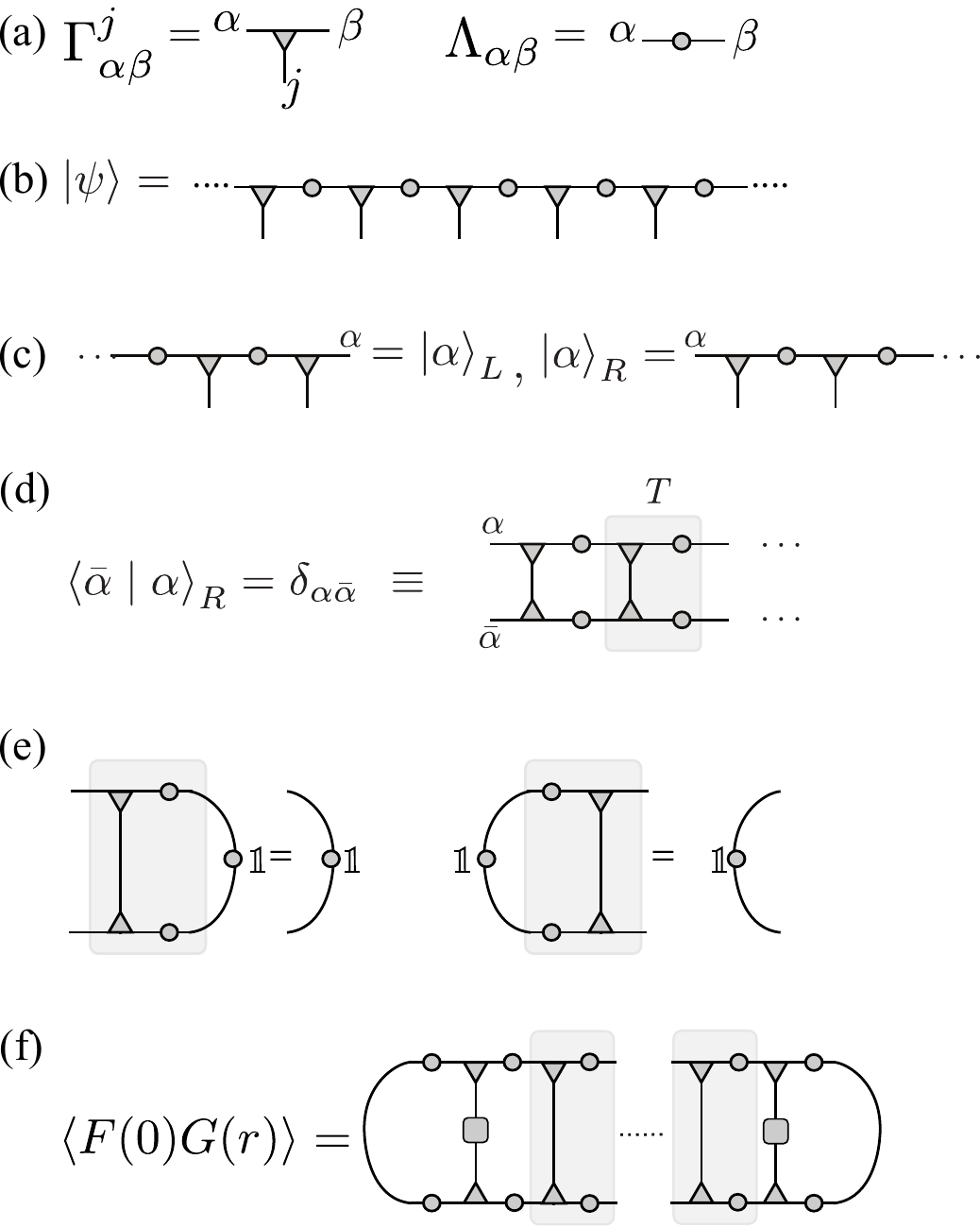}
	\caption{%
		Diagrammatic representation of (a) the tensors $\Gamma$ and $\Lambda$.  The horizontal lines represent the bond indices $\alpha \in \{ 1, \dots, \chi \}$ and the vertical lines the physical indices $j \in \{1, \dots, d\}$. 
		(b) An  MPS formed by the tensors $\Gamma$ and $\Lambda$.  Connected lines between tensors (or within a single tensor) denote summation over the corresponding indices.  (c) Definition of the right and left (Schmidt) basis states with respect to a partition on a bond with index $\alpha$.  (d) Condition for the MPS to be in the canonical form. The transfer matrix $T$ of Eq.~\eqref{eq:transfer_r} has been shaded. The upside-down triangles are the complex conjugate of the $\Gamma$ tensors.  (e) If the state is in canonical form, then the dominant right eigenvector of $T$ is the `identity matrix' with eigenvalue equal to $1$.  A similar condition applies for the left transfer matrix $\tilde{T}$. (f) The correlation function defined in Eq.~\eqref{corrfcn}. The squares correspond to the operators $F(0)$ and $G(r)$.
	}
	\label{fig:mps}
\end{figure}

\paragraph*{Canonical form.}
Without a loss of generality, we write the matrices $A^j$ as a product of $\chi_{j-1} \times \chi_j$ complex matrices $\Gamma^{j}$ and positive, real, square diagonal matrices $\Lambda$,
\begin{align}
	\ket{\Psi} = \sum_{j_1, \ldots, j_N}
		\begin{array}{r}
		\Gamma^{[1]j_1} \Lambda^{[1]} \Gamma^{[2]j_2} \Lambda^{[2]} \cdots \Lambda^{[N-1]} \Gamma^{[N]j_N}
			\,\\	\times\, | j_1, \ldots ,j_{N} \rangle
		\end{array}
	,  \label{eq:mpsGL}
\end{align}
which is pictorially illustrated in Figs.~\ref{fig:mps}(a) and \ref{fig:mps}(b).
A rank-$n$ tensor is represented by a symbol with $n$ protruding lines.
(For example, $\Gamma$, a rank-3 tensor, has three indices and is represented by a triangle with three lines protruding from it.)
Connecting the lines among tensors symbolizes a tensor contraction, i.e., summing over the relevant indices. In the following we will motivate the choice~\eqref{eq:mpsGL} for the  MPS form.

Equation~\eqref{eq:mpsGL} allows for many possible representation of the same wave function, as we can insert a resolution of the identity $\mathds{1} = X X^{-1}$ into any bond.  This freedom can be used to define a `canonical form' of the MPS, following Ref.~\onlinecite{Vidal-2003, Vidal-2007}. 
Any bond $n$ defines a bipartition of the system into sites $L = \{1,\dots,n\}$ and $R = \{n+1,\dots,N\}$ to the left and right of the bond. From the form of the MPS, we can define a set of $\chi_n$ wave functions $\ket{\alpha}^{[n]}_{L/R}$ to the left/right of the bond [see Fig.~\ref{fig:mps}(c)] such that state takes the form
\begin{align}
	\ket{\psi} &= \sum_{\alpha=1}^\chi \Lambda^{[n]}_\alpha \ket{\alpha}^{[n]}_L \otimes \ket{\alpha}^{[n]}_R,
		\quad \ket{\alpha}^{[n]}_{R/L} \in \mathcal{H}_{R/L}.
	\label{eq:schmidt}
\end{align}
The MPS representation $\{\Gamma^{[1]}, \Lambda^{[1]}, \dots, \Gamma^{[N]}\}$ is in canonical form if:
\emph{For every bond, the set of wave functions $\ket{\alpha}^{[n]}_{L/R}$ along with $\Lambda^{[n]}$ form a Schmidt decomposition of $\Psi$.} In other words we must have $\braket{\bar\alpha | \alpha}^{[n]}_L = \delta_{\bar\alpha\alpha}$ and $\braket{\bar\alpha | \alpha}^{[n]}_R = \delta_{\bar\alpha\alpha}$, along with $\sum (\Lambda^{[n]}_\alpha)^2 = 1$ on every bond.
For finite systems, a generic MPS can be transformed into canonical form by successively orthogonalizing the bonds starting from either the left or right end of the chain. \cite{Schollwoeck11}

\paragraph*{Infinite matrix product states.}

In this paper we are most interested in infinite chains, $N\rightarrow \infty$. In this case, translational invariance is restored and the set of matrices on any given site becomes the same, that is  $\Gamma^{[n]j} = \Gamma^{j}$ and $\Lambda^{[n]}=\Lambda$ for all integers $n$.
To check if the iMPS is in canonical form, we need to compute the overlaps $\ev{\bar{\alpha}|\alpha}_R$,
which would appear to require an infinite tensor contraction. But, we can use the translation invariance to proceed inductively.
For infinite MPS, we can conveniently express the orthogonality condition (i.e., canonical form) in terms of a \emph{transfer matrix} $T$ [illustrated in Fig.~\ref{fig:mps}(d)] defined as 
\begin{align}
	T_{\alpha \bar{\alpha}; \beta \bar{\beta} } &= \sum_{j}
		\Gamma^j_{\alpha \beta} \, \big(\Gamma^j _{ \bar{\alpha} \bar{\beta} }\big)^{\ast}
		\Lambda _{\beta} \Lambda _{\bar{\beta}} ,
	\label{eq:transfer_r}
\end{align}
where `$^\ast$'~denotes complex conjugation (and is pictorially represented by an upside-down triangle). The transfer matrix $T$ relates the overlaps defined on bond $n$ with overlaps defined on bond $n+1$.
Given that the right basis states $\ket{\beta}^{[n+1]}_R$ on bond $n+1$ are orthonormal, the states $\ket{\alpha}^{[n]}_R$ on bond $n$ will also be orthonormal if $T$ has a \emph{right} eigenvector $\delta _{\beta \bar{\beta}}(=\mathds{1})$ with eigenvalue $\eta=1$, as illustrated in Fig.~\ref{fig:mps}(e).
For the left set of states we define an analogous transfer matrix $\tilde{T}$,
\begin{align}
	\tilde{T}_{\alpha \bar{\alpha}; \beta \bar{\beta} } &= \sum_{j}
		\Lambda _{\alpha}\Lambda _{\bar{\alpha}} \,
		\Gamma ^{j}_{\alpha \beta} \, \big(\Gamma^j_{\bar{\alpha} \bar{\beta}}\big)^{\ast }
	\label{eq:transfer_l}
\end{align}
which must have a \emph{left} eigenvector $\delta _{\alpha \bar{\alpha}}$ with $\eta =1$. These eigenvector criteria are clearly necessary conditions for all bonds to be canonical; in fact, assuming in addition that $\eta = 1$ is the dominant eigenvalue, they are sufficient. An algorithm to explicitly transform an arbitrary infinite MPS to the canonical form is given in Ref.~\onlinecite{Orus-2008}. 

If the infinite MPS is not translational invariant with respect to a one-site unit cell, all the above can be simply generalized by considering a unit-cell of $L$ sites which repeats itself, e.g., in the case of $L=2$ the tensors are given by 
\begin{align}
	\begin{array}{c@{\,}c @{\quad\quad} c@{\,}c}
		\Gamma^{[2n]} &= \Gamma^{A},
		&\Lambda^{[2n]j} &= \Lambda^A, \\
		\Gamma^{[2n+1]} &= \Gamma^{B},
		&\Lambda^{[2n+1]} &= \Lambda^B,
	\end{array}
\end{align}
for $n\in\mathbb{Z}$.
Reviews of MPS's as well as the canonical form can be found in Refs.~\onlinecite{PerezGarcia-2007,Orus-2008,Vidal-2007}.

\paragraph*{Calculations of observables from an iMPS.}
If the MPS is given in canonical form, we can use the orthogonality of the Schmidt states to evaluate local expectation values  by contracting the tensors locally.\cite{Vidal-2007} Correlation functions can be obtain using the transfer matrix Eq.~\eqref{eq:transfer_r}. For this we evaluate $\braket{F(0)G(r)}$ of an iMPS.
Let $r > 0$, then
\begin{align}
\braket{F(0)G(r)} &= \Upsilon_L(F) T^{r-1} \Upsilon_R(G),\nonumber\\
\Upsilon_L(F)_{\alpha\bar{\alpha}} &= \sum_\gamma \Lambda_\gamma^2 \big(\Gamma^{i}_{\gamma\bar\alpha}\big)^\ast F^{ij} \Gamma^{j}_{\gamma\alpha} \Lambda_\alpha \Lambda_{\bar\alpha}	,\nonumber\\
\Upsilon_R(G)_{\beta\bar{\beta}} &= \sum_\gamma \Lambda_\gamma^2 \big(\Gamma^{i}_{\bar\beta\gamma}\big)^\ast G^{ij} \Gamma^{j}_{\beta\gamma}	.
\label{corrfcn}
\end{align}
$\Upsilon_{L/R}$ are the ``stubs'' which measures $F$ and $G$ locally, in between which we put $r-1$ copies of the transfer matrix $T$, see Fig.~\ref{fig:mps}(f) for a pictorial representation. Local observables $\braket{F(0)}$ can be obtained from the same expression, replacing $\Upsilon_R(G)$ and $T^0$ with identity operators. 
By generalizing the transfer matrix to include on-site operators, non-local order parameters can be obtained with the same approach
\begin{equation}
T^{R}_{\alpha \alpha ^{\prime };\beta \beta ^{\prime }}=\sum_{j}\left(\sum_{j^{\prime }}R _{jj^{\prime }}\Gamma_{j^{\prime },\alpha\beta}\right)\left(\Gamma _{j,\alpha ^{\prime
}\beta ^{\prime }}\right)^{\ast}\Lambda _{\beta }\Lambda _{\beta ^{\prime }}.
\label{nonlocalT}
\end{equation}
For example, calculating the correlation function with $F=G=S_z$ and  $R=e^{i\pi S^z}$, we obtain the ``string order'' Eq.~(\ref{eq:SO}).

The resulting correlation functions generically take the form of a sum of exponentials, with the slowest decaying exponential determined by the second largest (in terms of absolute value) eigenvalue 
$\epsilon_2$ of the transfer matrix. We define the correlation length of the MPS as
\begin{equation}
	\xi = - \frac{1}{\log{|\epsilon_2|}} \label{eq:corr},
\end{equation}
which is readily obtained using a sparse algorithm to find the eigenvalues of the transfer matrix. A degenerate largest eigenvalue indicates that the state is in a `cat state,' i.e., in a superposition of different superselection sectors, which can occur when there is spontaneous symmetry breaking.\cite{Perez09}

In systems with a conserved quantum number (e.g., the total $S_z$), one can calculate the correlation length for operators ($F,G$) corresponding to different sectors from the corresponding eigenvalues of the transfer matrix. In this paper we denote the two correlation lengths corresponding to operators which change the quantum numbers by $S_z=0$ (e.g.~$\braket{S^zS^z}$) as $\xi_0$ and $S_z = \pm 1$ (e.g.~$\braket{S^+S^-}$) as $\xi_1$.
The correlation length $\xi$ is given by the largest one, i.e., $\xi=\text{max}(\xi_0,\xi_1,...)$.


\subsection{Infinite Time Evolving Block Decimation (iTEBD)}
\begin{figure}[tb!]
  \begin{center}
    \includegraphics[width=8cm]{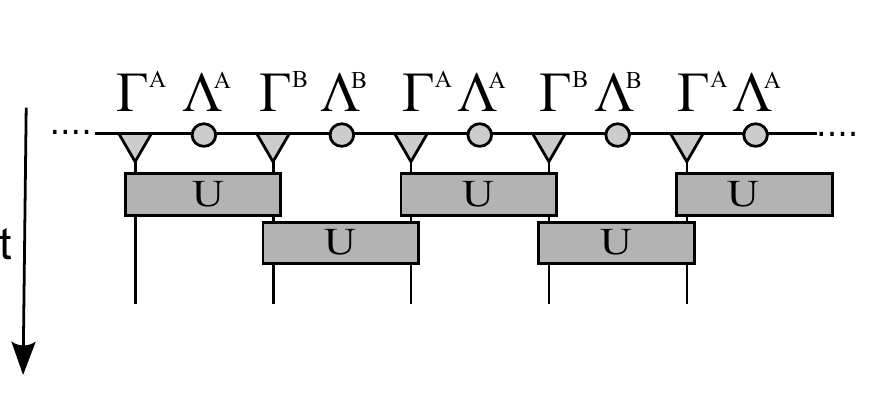}
  \end{center}
	\caption{In iTEBD each time step $\delta t$ of a time evolution is approximated using a Trotter-Suzuki decomposition, i.e., the time evolution operator is expressed as a product of unitary two-site operators.} 
  \label{fig:itebd}
\end{figure}
In the iTEBD algorithm, we are interested in evaluating the time evolution of a quantum state:
\begin{equation}
	\ket{\psi(t)} = U(t)\ket{\psi(0)}.
\end{equation}
The time evolution operator $U$ can either be $U(t) = \exp(-iHt)$ yielding a real time evolution, or an imaginary time evolution $U(\tau) = \exp(-H\tau)$. The latter is used to find ground states of the Hamiltonian $H$ through the relation
\begin{equation}
	\ket{\psi_G} = \lim_{\tau\rightarrow\infty} e^{-\tau H}\ket{\psi_0}.
\end{equation}
To achieve this, one makes use of the Trotter-Suzuki decomposition, which approximates the exponent of a sum of operators, with a product of exponents of the same operators. For example, the first order expansion reads
\begin{equation}
	e^{(A+B)\delta} = e^{A\delta}e^{B\delta} + \mathcal{O}(\delta^2).
	\label{eq:ST1}
\end{equation}
Here $A$ and $B$ are operators, and $\delta$ is a small parameter. The second order expansion similarly reads
\begin{equation}
	e^{(A+B)\delta} = e^{A\delta/2}e^{B\delta}e^{A\delta/2} + \mathcal{O}(\delta^3).
	\label{eq:ST2}
\end{equation}
To make use of these expressions, we assume that the Hamiltonian is a sum of two-site operators of the form $H=\sum_n h^{[n,n+1]}$ and decompose it as a sum 
\begin{align}
	H &= H_{\rm odd} + H_{\rm even} \notag\\
	&= \sum_{n\; {\rm odd}} h^{[n,n+1]} + \sum_{n\; {\rm even}} h^{[n,n+1]}. 
\end{align}
Each term $H_{\rm odd}$ and $H_{\rm even}$ consists of a sum of commuting operators.

We now divide the time into small time slices $\delta t\ll 1$ (the relevant time scale is in fact the inverse gap) 
and consider a time evolution operator $U(\delta t)$. Using, as an example, the first order decomposition~\eqref{eq:ST1}, the operator $U(\delta t)$ can be expanded into products of two-site unitary operators
\begin{equation}
	U(\delta t) \approx \left[\prod_{n\; {\rm odd}} U^{[n,n+1]}(\delta t)  \right]\left[\prod_{n\; {\rm even}} U^{[n,n+1]}(\delta t)  \right],
	\label{TimeEvol}
\end{equation}
where
\begin{eqnarray}
	U^{[n,n+1]}(\delta t)=e^{-i \, \delta t \, h^{[n,n+1]}}
\end{eqnarray}
This decomposition of the time evolution operator is shown pictorially in Fig.~\ref{fig:itebd}. One notices that even if the underlying system has a translation invariance of one site, the decomposition breaks this temporarily into a two site translation symmetry. Therefore, one needs to keep at least two sets of matrices $\Gamma^A,\Lambda^A$ and $\Gamma^B,\Lambda^B$. The successive application of these two-site unitary operators to an MPS is the main part of the algorithm.

\begin{figure}[tb!]
  \begin{center}
    \includegraphics[width=8cm]{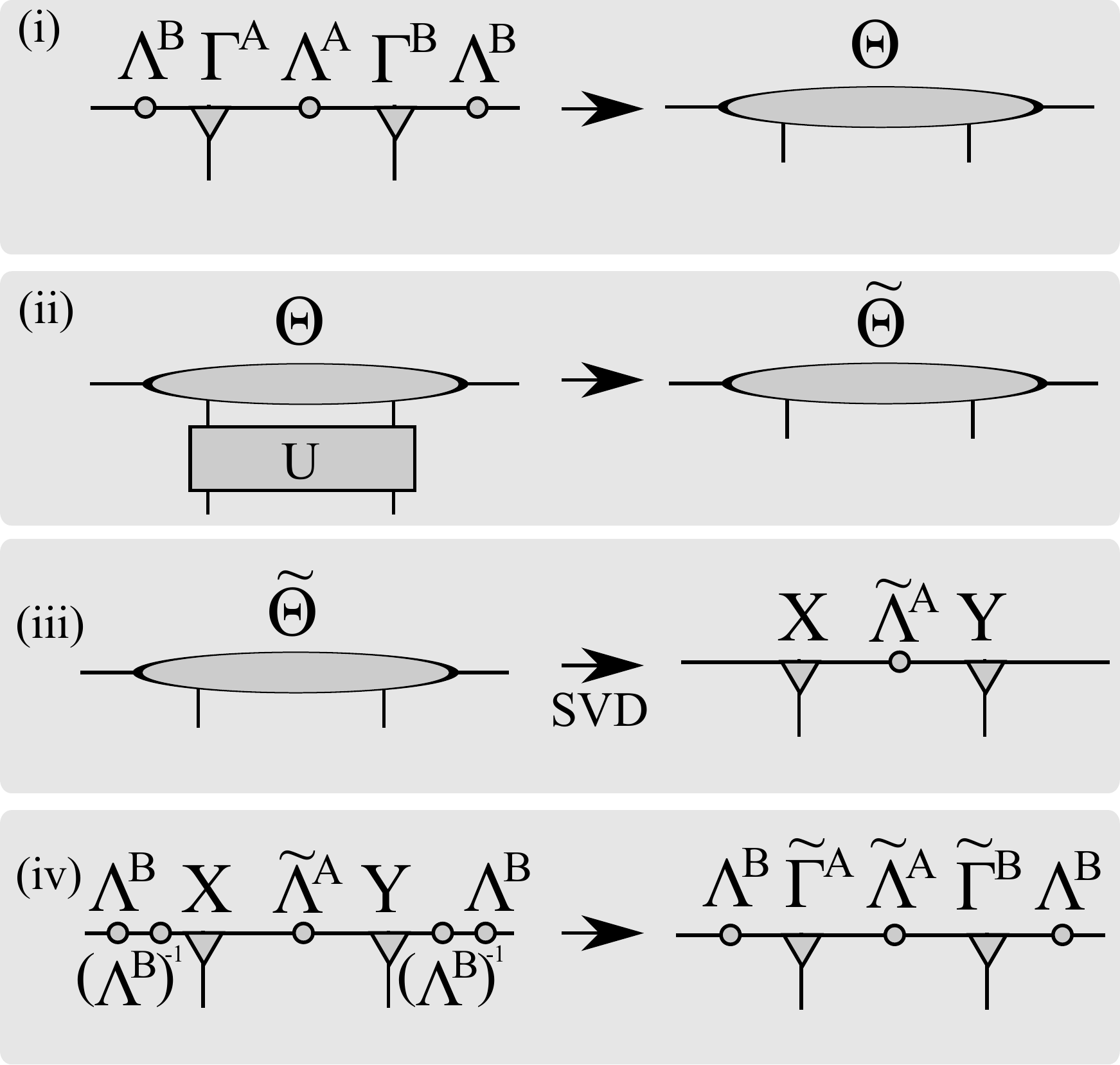}
  \end{center}
  \caption{The iTEBD update scheme for a unitary two-site transformation of a two-site unit cell MPS in canonical form (see text for details).} 
  \label{fig:update}
\end{figure}

\paragraph*{Local unitary updates of an MPS.}
One of the advantages of the MPS representation is that local transformations can be performed efficiently. Moreover, the canonical form discussed above is preserved if the transformations are unitary.\cite{Vidal-2003} 

A one-site unitary $U$ simply transforms the tensors $\Gamma$ of the MPS
\begin{equation}
\tilde{\Gamma}^{j}_{\alpha\beta} = \sum_{j^{\prime}}U^{j}_{j^{\prime}}\Gamma^{j^{\prime}}_{\alpha\beta}.
\end{equation}
If we consider an infinite, translational invariant  MPS, this transformations implies the application of the unitary to \emph{all} equivalent sites simultaneously. In such case the entanglement of the wave-function is not affected and thus the values of $\Lambda$ do not change. 

The update procedure for a two-site unitary transformation acting on two neighboring sites is shown in Fig.~\ref{fig:update}.  We focus on an update of an $AB$ bond between two neighboring sites $n$ and $n+1$ for an MPS with a unit cell of size $N=2$. The inequivalent $BA$ bonds are updated similarly by simply exchanging $A$ and $B$. The generalization to an $L$-site unit cell is straightforward. We first find the wave function in the basis spanned by the left Schmidt states on bond $n-1:n$, the 1-site Hilbert space of sites $n$ and $n+1$, and the right Schmidt states on bond $n+1:n+2$, which together form an orthonormal basis 
$\{ \ket{\alpha_{n-1}}_L,  \ket{j_n}, \ket{k_{n+1}}, \ket{\gamma_{n+1}}_R \}$. Calling the wave function coefficients $\Theta$, the state is expressed as
\begin{equation}
\ket{\psi} = \sum_{\alpha,j,k,\gamma} \Theta^{jk}_{\alpha\gamma} \ket{\alpha_{n-1}}_L \ket{j_n} \ket{k_{n+1}} \ket{\gamma_{n+1}}_R .
\label{theta2}
\end{equation}
Using the definitions of $\ket{\alpha}_{L/R}$ shown in Fig.~\ref{fig:mps}(b), $\Theta$ is given by
\begin{equation}
\Theta_{\alpha\gamma}^{jk} = \sum_{\beta} \Lambda^{B}_{\alpha}\Gamma^{{A},j}_{\alpha\beta}\Lambda^{A}_{\beta} \Gamma^{{B},k}_{\beta\gamma}\Lambda^{B}_{\gamma}.\label{theta}
\end{equation}
Writing the wave function in this basis is useful because it is easy to apply the two-site unitary in step (ii) of the algorithm:
\begin{equation}
\tilde{\Theta}^{jk}_{\alpha\gamma} = \sum_{j^{\prime} k^{\prime}} U^{jk}_{j^{\prime}k^{\prime}} \Theta^{j^{\prime}k^{\prime}}_{\alpha\gamma}.
\end{equation}
Next we have to extract the new tensors $\tilde{\Gamma}^{{A}},\tilde{\Gamma}^{B}$ and $\tilde{\Lambda}^{A}$ from the transformed tensor $\tilde{\Theta}$ in a manner that preserves the canonical form. We first `reshape' the tensor $\tilde{\Theta}$ by combining indices to obtain a $d\chi\times d\chi$ dimensional matrix $\Theta_{j\alpha;k\gamma}$. Because the basis $\ket{\alpha_{n-1}}_L \ket{j_n}$ is orthonormal, as for the right, it is natural to decompose the matrix using the singular value decomposition (SVD) in step (iii) into
\begin{eqnarray}
\Theta_{j\alpha;k\gamma} = \sum_{\beta} X_{j\alpha;\beta} D_{\beta} Y_{\beta;k\gamma},
\label{thetaAfterSVD}
\end{eqnarray}
where $X,Y$ are isometries and $D$ is a diagonal matrix. The isometry $X$ relates the new Schmidt states $\ket{\beta_n}_{L}$ to the combined bases $\ket{\alpha_{n-1}}_{L} \ket{j_n}$. Analogously, the Schmidt states for the right site are obtained from the matrix $Y$. Thus the diagonal matrix $D$ contains precisely the Schmidt values of the transformed state, i.e., $\tilde{\Lambda}^{A}=D$. The new tensors  $\tilde{\Gamma}^{{A}},\tilde{\Gamma}^{B}$ can be extracted directly from the matrices $X,Y$ using the old matrices $\Lambda^{B}$ and the definition of $\Theta$ in Eq.~(\ref{theta}). In particular we obtain the new tensors in step (iv) by 
\begin{eqnarray}
\tilde{\Gamma}^{{A},j}_{\alpha\beta}&=& (\Lambda^{B})^{-1}_{\alpha} X_{j\alpha;\beta}\\
\tilde{\Gamma}^{{B},j}_{\beta\gamma}&=& Y_{\beta;k\gamma}(\Lambda^{B})^{-1}_{\gamma}.
\end{eqnarray}
After the update, the new MPS is still in the canonical form. Note that as in the one-site update, if we apply the algorithm to an MPS, the update is performed simultaneously to all matrices at equivalent bonds.  Thus the iTEBD algorithms exploits the translational invariance of the systems by effectively performing  an infinite number of parallel updates at each step.  

The entanglement at the bond $n,n+1$ has, in the update, changed and the bond dimension increased to $d\chi$. Thus the amount of information in the wave function grows exponentially if we successively apply unitaries to the state. To overcome this problem, we perform an approximation by fixing the maximal number of Schmidt terms to $\chi$. After each step, only the $\chi$ most important states are kept, i.e., if we order the Schmidt states according to their size we simply truncate the range of the index $\beta$ in~\eqref{thetaAfterSVD} to be $1\dots \chi$. This approximation limits the dimension of the MPS and the tensors $\Gamma$ have at most a dimension of $d\times\chi\times\chi$. Given that the truncated weight is small, the normalization conditions for the canonical form will be fulfilled to a good approximation.
In order to keep the wave function normalized, one should divide by the norm after the truncation, i.e., divide by $\mathcal{N} = \sqrt{\sum_{i,j,\alpha,\gamma} \big|\Theta^{ij}_{\alpha\gamma}\big|^2}$.

If we perform an imaginary time evolution of the state, the operator $U$ is not unitary and thus it does not conserve the canonical form. It turns out, however, that the successive Schmidt decompositions assure a good approximation as long as the time steps are chosen small enough.\cite{Orus-2008} One way to obtain very accurate results is to decrease the size of the time steps successively. \cite{Vidal-2007}

The simulation cost of this algorithm scales as $d^3\chi^3$ and the most time consuming part of the algorithm is the SVD in step (iii). If the Hamiltonian has symmetries, we can considerably accelerate this step by explicitly conserving the resulting constants of motion. The anisotropic spin model we study has for example a global $U(1)$ symmetry and conserves the total magnetization. Thus the matrix $\Theta_{i\alpha;j\gamma}$ has a block-diagonal form and the SVD can be performed in each block individually,  yielding a considerable speed up. 
See Refs.~\onlinecite{Vidal2010,Vidal2011,Singh2012} for the details of the implementation of symmetries into the algorithm. Numerically, the algorithm can become unstable when the values of $\Lambda$ become very small since the matrix has to be inverted in order to extract the new tensors in step (iv) of the algorithm. This problem can be avoided applying a slightly modified version of this algorithm as introduced by Hastings in Ref.~\onlinecite{Hastings09}.

\subsection{Matrix-Product Operators}
\begin{figure}[tb!]
  \begin{center}
    \includegraphics[width=8cm]{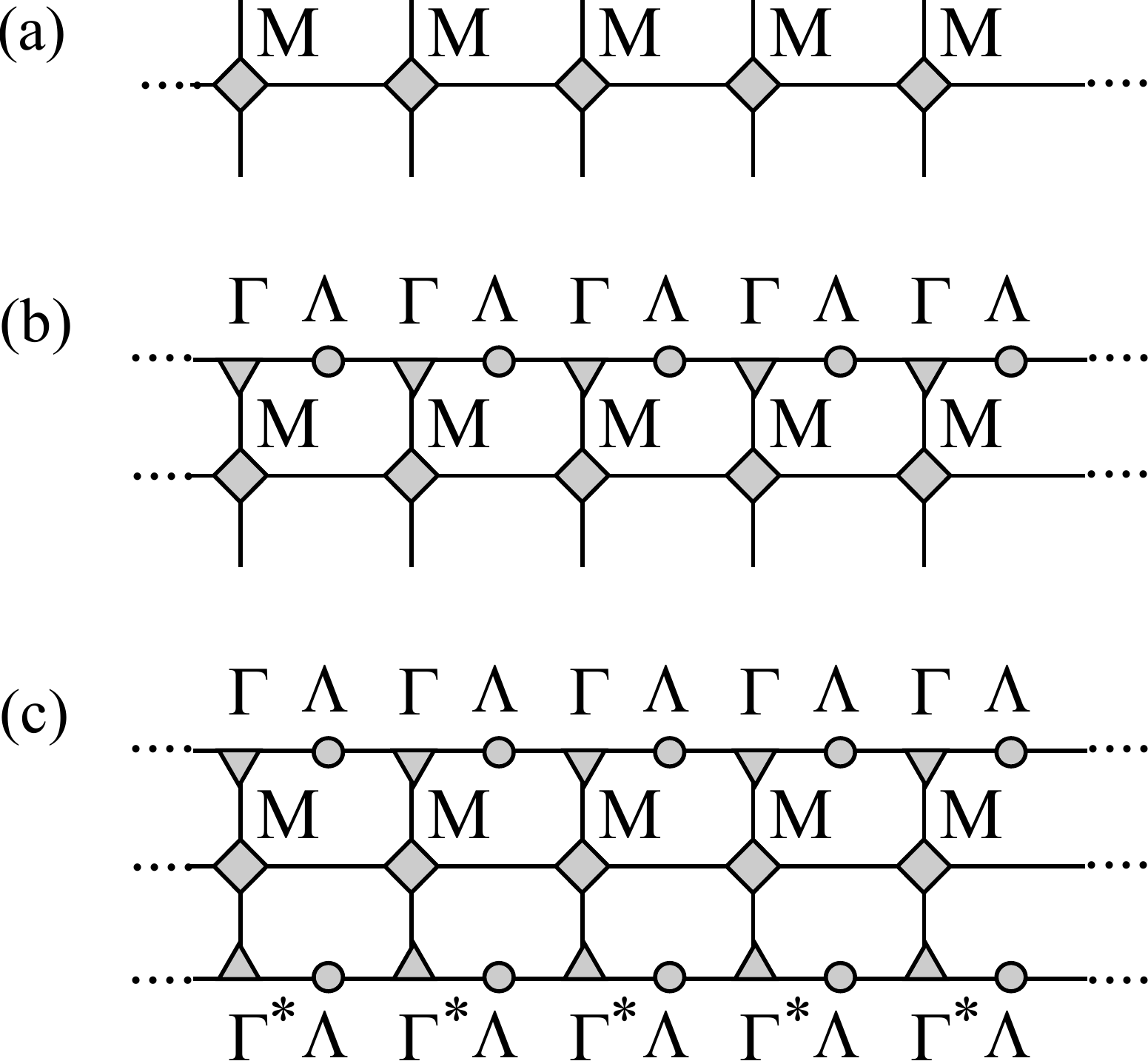}
  \end{center}
  \caption{(a) An operator $O$ acting on an entire chain expressed as a matrix product operator. (b) A matrix product operator acting on a matrix product state $O|\psi\rangle$. (c) The expectation value $\langle\psi|O|\psi\rangle$ expressed in an MPO form.} 
  \label{fig:mpo}
\end{figure}

The iDMRG algorithm explained in the next section relies on expressing the Hamiltonian of the system in terms of matrix product operator (MPO). An MPO is a natural generalization of an MPS to the space of operators. An operator in an MPO form, acting on a chain with $L$ sites, is given by   
\begin{eqnarray}
	O = \!\!\!\sum_{
		\begin{smallmatrix} j_1, \ldots, j_L\\ j_1^{\prime}, \ldots, j_L^{\prime} \end{smallmatrix}} \!\!\!
		\begin{array}{r}
			\\
			\vec{v}_\text{left} M^{[1]j_1j_1^{\prime}} M^{[2]j_2j_2^{\prime}} \ldots M^{[L]j_Lj_L^{\prime}} \vec{v}_\text{right}	\\ 
			\times \ket{j_1, \ldots ,j_L} \bra{j_1^{\prime}, \ldots ,j_L^{\prime}} \;,
		\end{array}
	\label{eq:mpo}
\end{eqnarray}
where $M^{j_nj_n',}$ are $D \times D$ matrices, and $|j_n\rangle$, $|j_n^{\prime}\rangle$ represent local states at site $n$, as before. At the boundaries we initiate and terminate the MPO by the vectors $\vec{v}_\text{left} $ and $ \vec{v}_\text{right}$.

A pictorial representation of an MPO is given in Fig.~\ref{fig:mpo}(a).
The notation is very similar to the one for an MPS: the horizontal line corresponds to the indices of the virtual dimension and the vertical lines represent the physical states $|j_n\rangle$ (bottom) and $\langle j_n^{\prime}|$ top.
The advantage of the MPO is that it can be applied efficiently to a matrix product state as shown in  Fig.~\ref{fig:mpo}(b). 
All local Hamiltonians with only short range interactions can be represented using an MPO of a small dimension $D$. Let us consider, for example, the MPO of the anisotropic Heisenberg model~\eqref{XXZ} in the presence of an on-site anisotropy. Expressed as a tensor product, the Hamiltonian takes the following form:
\begin{eqnarray} 
	H &=& S^x \otimes S^x \otimes \mathds{1} \otimes\dots  \otimes\mathds{1} + \mathds{1} \otimes S^x \otimes S^x \otimes\dots  \otimes\mathds{1} +\dots\nonumber\\
	&+& S^y \otimes S^y \otimes \mathds{1} \otimes\dots  \otimes\mathds{1} + \mathds{1} \otimes S^y \otimes S^y \otimes\dots  \otimes\mathds{1} +\dots\nonumber\\
	 &+&  \Delta S^z \otimes S^z \otimes \mathds{1} \otimes\dots  \otimes\mathds{1} + \dots \nonumber\\
	  &+&  [D_2 (S^z)^2+D_4(S^z)^4] \otimes \mathds{1}  \otimes \mathds{1} \otimes \dots  \otimes\mathds{1} +\dots
	\label{eq:hamil}
\end{eqnarray}
The corresponding \emph{exact} MPO has a dimension $D=5$ and is given by
\begin{align}
M^{[i]}=\begin{pmatrix}
	\mathds{1} & 0 & 0 & 0 & 0  \\
	S^x & 0 & 0 & 0 & 0  \\
	S^y & 0 & 0 & 0 & 0  \\
	\Delta S^z & 0 & 0 & 0 & 0  \\
  {\scriptstyle D_2 (S^z)^2+D_4 (S^z)^4} & S^x & S^y & S^z &  \mathds{1}
	\end{pmatrix} ,
\end{align}
with 
\begin{align}
\label{eq:vvec}
	\vec{v}_\textrm{left} = \begin{pmatrix} 0, & 0, & 0, & 0, & 1 \end{pmatrix} ,\quad
	\vec{v}_\textrm{right} = \begin{pmatrix} 1, & 0, & 0, & 0, & 0 \end{pmatrix}^T .
\end{align}
By multiplying the matrices (and taking tensor products of the operators), one can easily see that the product of the matrices does in fact yield the Hamiltonian~\eqref{eq:hamil}. Further details of the MPO form of operators can be found in Refs.~\onlinecite{Schollwoeck11,McCulloh-2008}.

\subsection{Infinite Density Matrix Renormalization Group (iDMRG)}
\begin{figure}[tb!]
  \begin{center}
    \includegraphics[width=8cm]{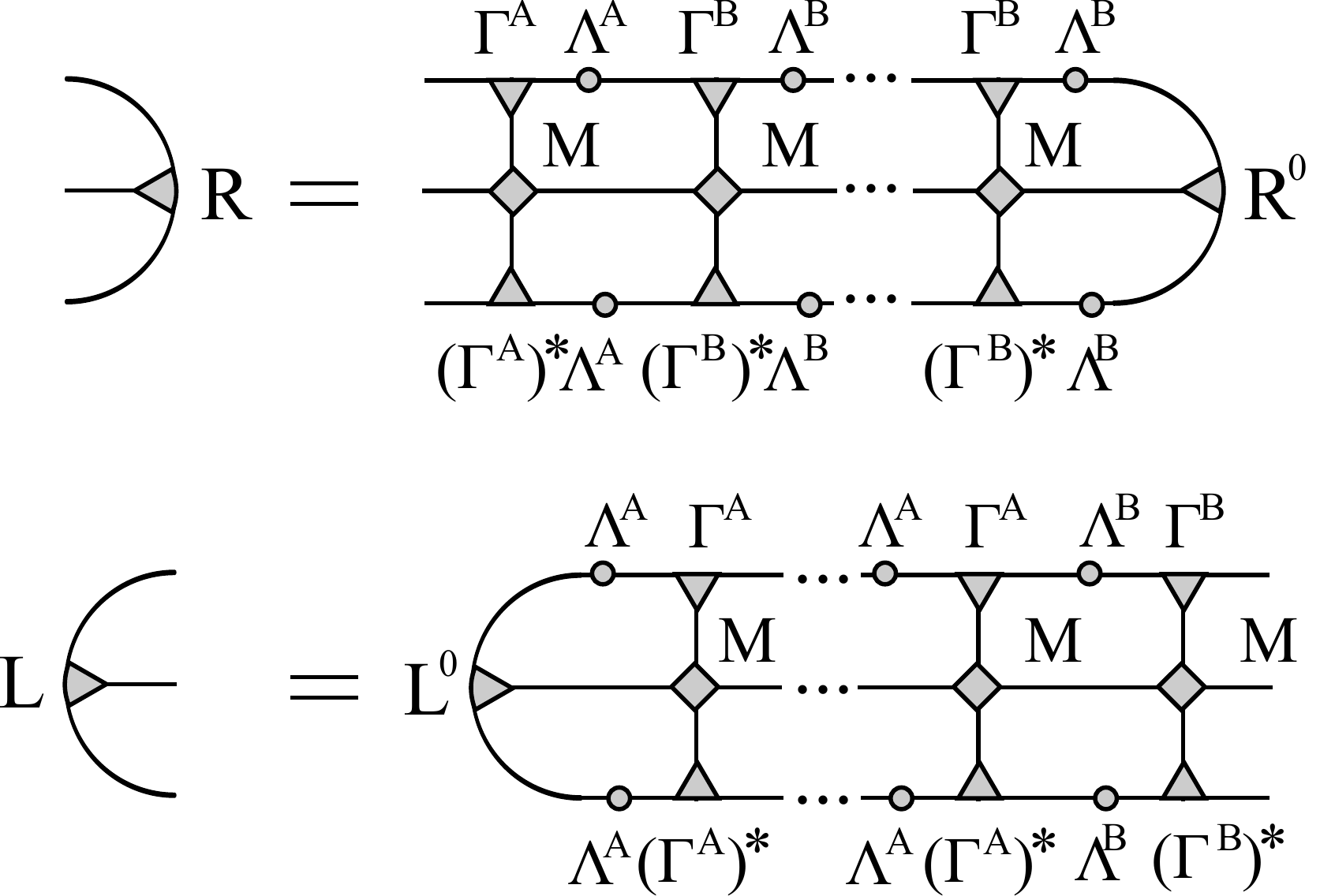}
  \end{center}
	\caption{Pictorial representation of a contraction of the left and right environments. The boundaries are initiated by the tensors $R^0_{\alpha,\bar{\alpha},a}=\delta_{\alpha, \bar{\alpha}} \vec{v}_{\text{right}; a}$ and  $L^0_{\alpha, \bar{\alpha},a}=\delta_{\alpha, \bar{\alpha}} \vec{v}_{\text{left}; a}$.}
  \label{environment}
\end{figure}

\begin{figure}[tb!]
  \begin{center}
    \includegraphics[width=8cm]{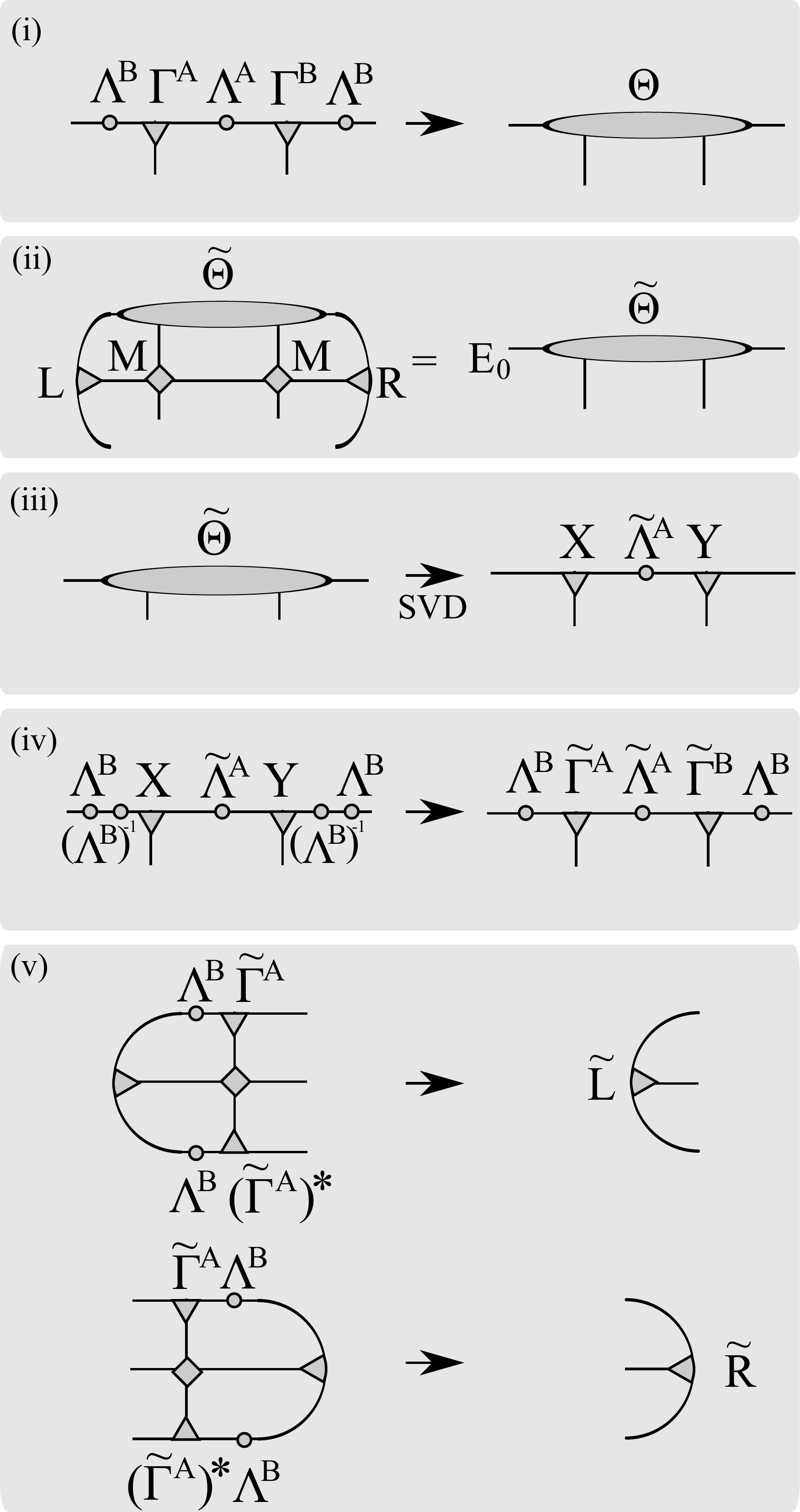}
  \end{center}
  \caption{A pictorial representation of an iDRMG iteration step update. See text for details.}
  \label{fig:idmrg}
\end{figure}

We now discuss the infinite Density Matrix Renormalization Group (iDMRG) algorithm. Unlike iTEBD, the iDMRG is a variational approach to optimizing the MPS, but the algorithms have many steps in common. One advantage of the iDMRG is that it does not rely on a Trotter-Suzuki decomposition of the Hamiltonian and thus applies to systems with longer range interactions. We assume only that the Hamiltonian has been written as an MPO. Secondly, the convergence of the iDMRG method to the ground state is in practice much faster. This is particularly the case if the gap above the ground state is small and the correlation length is long.

The schematic idea for the iDMRG algorithm is as follows (see Fig.~\ref{fig:idmrg}). Like in iTEBD, the state at each step is represented by an MPS.  We variationally optimize pairs of neighboring sites to minimize the ground state energy $\me{\psi}{H}{\psi}$, while keeping the rest of the chain fixed.  To do so, at each step we represent the initial wave function $\ket{\psi}$ using the two site tensor $\Theta^{jk}_{\alpha\gamma}$ (as previously defined in the iTEBD section), project the Hamiltonian into the space spanned by the basis set $\ket{\alpha j k \beta}$, and use an iterative algorithm (e.g.~Lanczos) to lower the energy.  Repeating this step for each pair, the wave function converges to the ground state. For simplicity, only the details of the algorithm with a unit cell of two sites, $A$ and $B$, will be described below.

\paragraph*{Two-site update algorithm.}
	We start by describing  the update of an $AB$ bond between two neighboring sites $n$ and $n+1$ (the update on a BA bond can be performed analogously by exchanging the role of A and B), and return later to the initialization procedure.
Step (i) is identical to the first step in the iTEBD method; we contract the tensors for two neighboring sites to obtain the initial wave function $\Theta^{jk}_{\alpha \gamma}$.
The orthonormal basis $\ket{\alpha j \beta k}$ spans the variational space $\ket{\tilde{\psi}} = \tilde{\Theta}^{jk}_{\alpha \gamma} \ket{\alpha j \beta k}$ of the update, in which we must minimize the energy $E = \langle \tilde{\psi}|H|\tilde{\psi}\rangle$ in order to determine the optimal $\tilde{\Theta}$.
Because $H$ is written as an infinite MPO, it appears at first that to evaluate the energy we will have to contract an infinite number of tensors starting from left and right infinity, as illustrated in Fig.~\ref{fig:mpo}(c). For the sake of induction, however, suppose we have already done this contraction on the left through bond $n-1:n$, and on the right through bond $n+1:n+2$.
As illustrated in Fig.~\ref{environment}, the result of these contractions can be summarized in two three leg tensors we call the left and right ``environments.''
The left environment $L_{\alpha \bar{\alpha}, a}$ has three indices: the MPO index $a$, and the indices $\alpha, \bar{\alpha}$ corresponding to the bond indices of $\ket{\tilde{\psi}}$ and $\bra{\tilde{\psi}}$.
Likewise, on the right we have $R_{\gamma \bar{\gamma}, c}$.  
Each bond of the system has a similarly defined environment; for a unit cell of two, we have in total $\{ L^A, L^B \}, \{ R^A , R^B \}$.
These environments are nothing other than the MPO for the Hamiltonian projected into the space of left and right Schmidt states about each bond.

With the environment in hand, we can project the Hamiltonian into the orthonormal basis $\ket{\alpha j \gamma k}$; to minimize the energy of $\Theta$ we find the ground state of the $\chi^2 d^2 \times \chi^2 d^2$ ``Hamiltonian'':
\begin{equation}
H_{ \alpha j k \gamma; \bar{\alpha} \bar{j} \bar{k} \bar{\gamma}} = \sum_{\gamma,a,b,c} L^B_{\alpha \bar{\alpha},a} M^{j,\bar{j}}_{ab} M^{k,\bar{k}}_{bc} R^B_{\gamma  \bar{\gamma},c}.
\end{equation}
To find this ground state, we use an iterative procedure such as Lanczos or Jacobi-Davidson at a cost of $\chi^3 D d^2$ per multiplication, as illustrated in step (ii), and obtain an improved guess for the wave function $\tilde{\Theta}$ and energy $E_0$.
By using the initial wave function $\Theta$ as the starting vector for the minimization procedure, convergence is typically reached with only a couple of steps.
This can be compared to the iTEBD optimization where we obtain a new wave-function $\tilde{\Theta}$ after applying the imaginary time-evolution operator.
As with iTEBD, the bond dimension grows as $\chi \to d \chi$, which we must truncate using SVD, shown in step (iii).
It is important that the left and right Schmidt basis about any bond remain orthogonal, because we assume $\ket{\alpha j \beta k}$ is an orthogonal basis at each step. Assuming this was the case on bonds of type $B$, the isometry properties of the SVD matrices $X$ and $Y$ imply that the orthogonality condition holds for the updated Schmidt states defined about the central bond $A$, and hence will remain so throughout the simulation.
At this point, we have improved guesses for the matrices $\tilde{\Gamma}^{A/B}, \tilde{\Lambda}^{A}$ in step (iv). 

The last step is to update the environment. At a minimum, we must update the environments on the bond which we just optimized by simply multiplying new tensors to the left and right as shown in Fig.~\ref{fig:idmrg}(v):
\begin{align}
\tilde{L}^A_{\beta \bar{\beta}, b} &= L^B_{\alpha \bar{\alpha}, a} \Lambda^B_\alpha \tilde{\Gamma}^{A}_{\alpha \beta j}  M^{j,\bar{j}}_{ab} \Lambda^B_{\bar{\alpha}} \tilde{\Gamma}^{A}_{\bar{\alpha} \bar{\beta} \bar{j}}, \\
\tilde{R}^A_{\beta \bar{\beta}, b} &= R^B_{\gamma \bar{\gamma}, a}  \tilde{\Gamma}^{B}_{\beta \gamma k} \Lambda^B_\gamma  M^{k,\bar{k}}_{ab}  \tilde{\Gamma}^{A}_{\bar{\gamma} \bar{\beta} \bar{k}} \Lambda^B_{\bar{\gamma}}.
\end{align}
This concludes the update on bond $AB$ and we move over by one site, exchanging the roles of $A$ and $B$, and repeat until convergence is reached.

\paragraph*{Initializing the environment.}
We now return to the problem of initializing the algorithm.
The initial MPS can be arbitrary (though it should be in canonical form).
A fine choice is a $\chi = 1$ tensor product state which either preserves or breaks the symmetries as desired.
To form the initial environment, we suppose when computing the left/right environment that $\hat{H}$ is zero to the left/right of the bond, which is captured by tensors of the form
\begin{align}
\label{eq:envinit}
R^{[n]}_{\alpha, \bar{\alpha},a}&=\delta_{\alpha, \bar{\alpha}} \vec{v}_{\textrm{right}; a},\\
L^{[n]}_{\alpha, \bar{\alpha},a}&=\delta_{\alpha, \bar{\alpha}} \vec{v}_{\textrm{left}; a},
\end{align}
where the $\vec{v}_{\textrm{left}/\textrm{right}}$ are as in Eq. \eqref{eq:mpo}.
Referring to Eq. \eqref{eq:vvec} as an example, recall that $\vec{v}_{\textrm{right}}$ specifies the MPO index such that no further operators will be inserted to its right; likewise, $\vec{v}_{\textrm{left}}$ indicates no operators have been inserted to its left.
Because all terms in the Hamiltonian then act as the identity to the left/right of the bond, the orthogonality of the Schmidt vectors implies that projecting the identity operator into the left/right Schmidt basis trivially gives $\delta_{\alpha, \bar{\alpha}}$.
When symmetry breaking is expected it is helpful to further initialize the environments by repeatedly performing the iDMRG update \emph{without} performing the Lanczos optimization, which builds up environments using the initial symmetry broken MPS.

\paragraph*{Ground state energy from iDMRG.}
One subtlety of the above prescription lies in the interpretation of the energy $E_\text{GS}$ obtained during the diagonalization step.
Is it the (infinite) energy of the infinite system?
Using the initialization procedure just outlined, the Lanczos energy $E_\text{GS}$ after the first step is the energy of the two-site problem.
While we motivated the environments as representing infinite half chains, it is more accurate to assign them a length of 0 after the initialization procedure, and at each optimization step the length of the left/right environment about the central bond increases because a site has been appended.
Keeping track of the length $\ell_{R/L}$ of each environment (for a unit cell of two, each grows on alternate steps), we see that the energy $E_\text{GS}$ corresponds to a system of size $\ell = \ell_L + 2 + \ell_R$.
By monitoring the change in $E_\text{GS}$ with increased $\ell$, we can extract the energy per site. This is convenient for problems in which there is no few-site Hamiltonian with which to evaluate the energy.

As for the iTEBD algorithm, we can achieve a considerable speed-up by using the symmetries of the Hamiltonian, which requires assigning quantum numbers to the tensors of the MPO in addition to the MPS.

\subsection{Finite entanglement scaling}\label{sec:fes}

An advantage of the infinite system methods introduced above is that no artifacts from the boundary appear. On the other hand, finite size effects can be very useful for performing a scaling analysis. In this section we show that critical properties of the system can be extracted by performing a ``finite-entanglement scaling'' in the infinite systems. This means, one can perform simulations with different bond-dimensions $\chi$ at a critical point and use the induced {\it finite} correlation length $\xi_{\chi}$ as a scaling variable analogous to a finite system size. 

To motivate this notion, consider the entanglement entropy $S_E$, which for an infinite system diverges logarithmically as a function of the correlation length as criticality is approached.~\cite{Calabrese} In an MPS, however, $S_E$ is bounded by $S_E\le\log\chi$, and an infinite $\chi$ is needed to accurately represent critical states. Clearly we cannot perform simulations with an infinite $\chi$, raising the question: what happens if we nevertheless optimize a finite dimensional MPS for a critical system? This question has been addressed by a series of papers.~\cite{Tagliacozzo,Pollmann09,Pirvu-2012}
It turns out that  simulating critical systems using finite $\chi$ cuts off long distance correlations a finite length $\xi_{\chi}$.
If we define the correlation length of the MPS $\xi$ to be the length obtained from the second largest eigenvalue of the transfer matrix, as define in Eq.~(\ref{eq:corr}),
then \emph{at criticality} the correlation length of the MPS scales as 
\begin{equation}
\xi \propto \chi^{\kappa}
\label{fet}
\end{equation}
where $\kappa \approx \frac{1}{c}\frac{6}{\sqrt{\frac{12}{c}}+1}$.~\cite{Pollmann09}
Because $\chi$ introduces a length scale in a universal way, we can \emph{define} the `finite entanglement length' by $\xi_{\chi} \equiv C\chi^{\kappa}$, where $C$ is independent of $\chi$, and extract various quantities of interest using a finite $\xi_{\chi}$ scaling analysis, or ``finite entanglement scaling.''
In an infinite system at criticality, the scaling relations are generally obtained from the analogous scaling relations in a finite size system  by replacing the finite length $L$ by $\xi_{\chi}$.  
For example, for a critical point with central charge $c$, the entanglement entropy $S_E$ between  two halves of a finite system of length $L$ scales as $S_E = \frac{c}{6} \log(L/a) + s_0$, with $a$ the lattice spacing and $s_0$ a non-universal constant.
If we instead measure $S_E$ for an \emph{infinite} system, but with finite $\chi$, we can substitute $L \to \xi_\chi$, 
\begin{equation}
S_E=\frac{c}{6}\log(\xi_{\chi}/a) + s_0'.
\label{tc}
\end{equation}
The additive constant is again non-universal, and unrelated to $s_0$. 
One should note, though, that while $\xi_\chi$ and $L$ have the same scaling dimension (i.e., that of length), the actual scaling functions are not guaranteed to be the same (for example, $s_0 \neq s_0'$ in the above).

Another useful quantity at criticality for systems with a $U(1)$ symmetry is the `stiffness,' here parameterized as the Luttinger parameter $K$.~\cite{Giamarchibook} For Hamiltonians that conserve the total magnetization, it can be obtained from the scaling of bipartite spin fluctuations of a half chain
\begin{equation}
F=\langle (S^z_L)^2\rangle - \langle S^z_L\rangle ^2,
\end{equation}
where $S^z_L$ is the $z$-component of the total spin to the left of a cut (for example, the total magnetization of the sites $i < 0$). The spin fluctuations satisfy~\cite{Song10,Song11}
\begin{equation}
F=\frac{K}{2\pi^2}\log(\xi_{\chi}/a) + \text{const},
\label{Kc}
\end{equation}
allowing us to extract $K$ by measuring the scaling of $F$ with increased $\chi$.

\begin{figure}[tb!]
  \begin{center}
    \includegraphics[width=8cm]{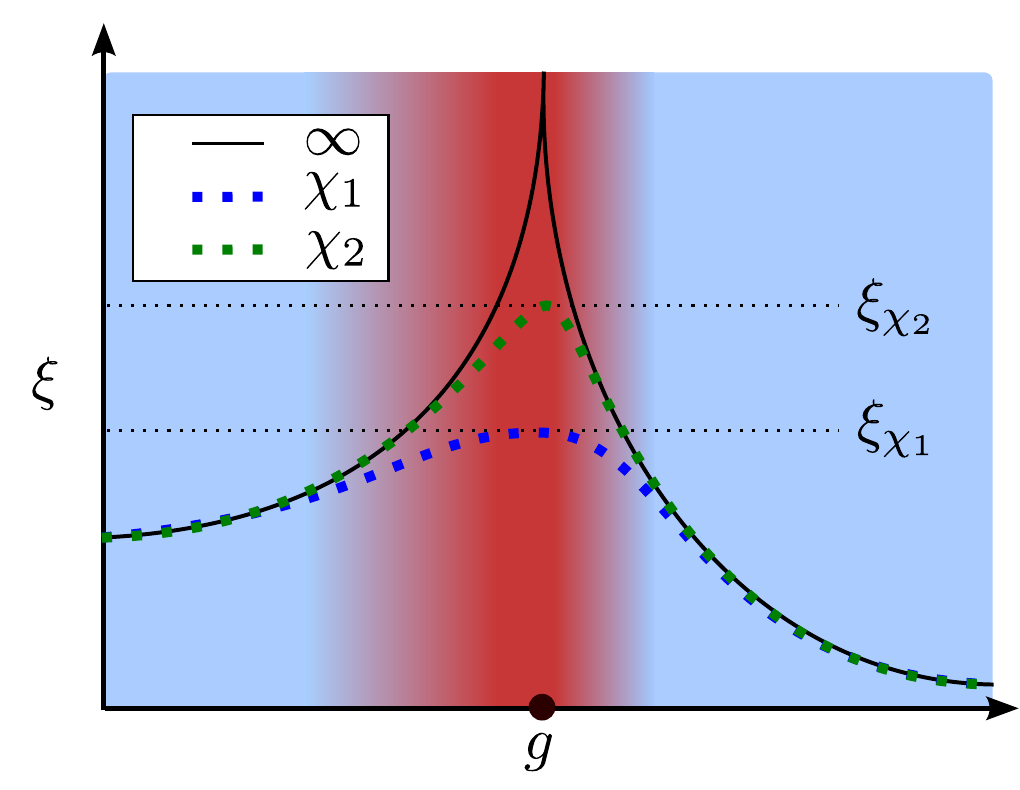}
  \end{center}
  \caption{(Color online) A schematic plot illustrating the idea of finite-entanglement scaling. The black solid line shows the exact correlation length $\xi_{\text{phys}}$ of the Hamiltonian, which diverges at the critical point. The dotted lines show the correlation lengths $\xi$ of the optimized iMPS at a finite $\chi_2 >\chi_1$. The horizontal dashed lines are the correlation lengths $\xi_{\chi_1(\chi_2)}$ from Eq.~(\ref{fet}), which are induced by the finite-entanglement cut-off at the critical point. The color shaded background indicates the two different regimes: Blue is the regime in which the iMPS is converged to the exact ground state using bond dimensions $\chi_1,\chi_2$ and red is the scaling regime which shrinks with increasing $\chi$  (see main text for further details).
}
  \label{fig:fes}
\end{figure}

The above discussion holds at criticality. Next we discuss the situation in the vicinity of a critical point, where the physical correlation length $\xi_{\text{phys}}$ is finite but much larger than the correlation length induced by finite $\chi$, i.e. $\xi_\chi \ll \xi_{\text{phys}} $.
For our purposes, we can define $\xi_{\text{phys}}$ by the MPS correlation length $\xi$ in the limit $\chi \to \infty$, where the MPS represents the true ground state.
In this regime the MPS is still cutoff by $\xi_\chi$, rather than the true correlation length $\xi_{\text{phys}}$,
so the finite-entanglement scaling relations Eqs.~\eqref{tc} and~\eqref{Kc} can still be used to obtain quantities like $c$ and $K$.
We will refer to this parameter range as the finite entanglement scaling region and it is shown as the red area in the schematic plot in Fig.~\ref{fig:fes}.  

Further away from criticality, as $\xi_{\text{phys}}  \ll \xi_{\chi} $, we can fully converge the MPS and the state `knows' it is not at the critical point. In this regime $\xi \to \xi_{\text{phys}}$ is independent of $\chi$, and all other observables are converged in $\chi$. If we measure the critical quantities $c$ and $K$ using Eq.~\eqref{tc} and Eq.~\eqref{Kc}, we find they renormalize to zero. 

In summary, in the finite entanglement scaling regime, $1 \ll \xi_{\chi} \ll \xi_{\text{phys}}$, we expect Eqs.~\eqref{tc} and~\eqref{Kc} to produce the critical values $c$ and $K$, but as $\xi_{\text{phys}}  \ll \xi_{\chi} $, the MPS converges and we cross over to the true, non-critical values $c = K = 0$.

The crossover can be analyzed by the following general finite entanglement scaling form.
Let $g^i$ be a set of \emph{physical} parameters, such as the coupling constants or the physical dimension of the system, and let $F$ be a scaling observable.
Near criticality, $F(g)$ has a scaling form determined by the scaling dimensions of $F$ and $g^i$.
When the system is approximated by an MPS of finite $\chi$, a new length scale $\xi_\chi$ is introduced.
While $\chi$ itself has fixed scaling dimension, since $\xi_{\chi} \sim \chi^\kappa$, we find that it is numerically more stable to parameterize the effect of $\chi$ through the MPS correlation length $\xi$.
We then measure $F$ using the MPS, $F(g; \xi)$.

The finite entanglement scaling procedure asserts that the usual scaling theory still applies to $F(g; \xi)$, with the addition of a single parameter of mass dimension $[\xi] = -1$.
As usual, the scaling hypothesis allows us to rescale $F, g, \xi$ in order to eliminate the dependence on one parameter. In the usual case in which there are no marginal operators, we can linearize the RG equations to determine the scaling dimensions $y_i$, $y_F$, and find
\begin{equation}
F(g^i, \xi) = e^{y_F \ell} F( e^{y_f \ell} g^i, e^{- \ell} \xi)
\end{equation}

Note that $\xi$ is in principle determined both by the physical correlation length $\xi_{\text{phys}}$ and the finite entanglement length $\xi_{\chi}$; as discussed, at criticality, $\xi = \xi_{\chi}$, while at infinite $\chi$, $\xi = \xi_{\text{phys}}$.
Regardless, $(g^i, \xi)$ remains a valid coordinate system for the parameters.

\section{\texorpdfstring {Determination of the $S=2$ phase diagram with \lowercase{i}DMRG} {Determination of the S=2 phase diagram with iDMRG}}	
\label{sec:QPT}
In this section we show how the numerical technique presented in the previous section was used to investigate the Hamiltonian Eq.~(\ref{Hamiltonian}) to establish the phase diagrams in Figs.~\ref{JzD2pd} and~\ref{D2D4pd}. We begin with a brief summary of the characteristics of the phases and the phase transitions in that model. We then show representative data used to obtain the spin-2 results of Sec.~\ref{sec:resultsummary} and discuss in more detail its conclusion. 

\subsection{Characterization of phases and phase transitions}
\paragraph*{Phases.}
The phase diagram of the model~\eqref{Hamiltonian} realizes several different phases, see for example Figs.~\ref{JzD2pd} and~\ref{D2D4pd}. For spin-2, the isotropic point, i.e., the Heisenberg model, lies in the trivial EH phase in the sense that it does not break any symmetry and contains the product state~\eqref{eq:productstate}. 

The FM and AFM phases are magnetically ordered with a nonzero magnetization on each site $\langle S_n^z\rangle\neq 0$. The FM phase has nonzero total magnetization $\langle S_n^z\rangle =\langle S_{n+1}^z\rangle$, while the AFM phase has zero total magnetization $\langle S_n^z\rangle =-\langle S_{n+1}^z\rangle$. All the other phases are nonmagnetic $\langle S_n^z\rangle=0$. 

The OH phase is an SPTP \cite{Berg-2008,Gu-2009,Pollmann-2010,Pollmann12,ChenGu-2011,ChenGu-2011-2,Schuch-2011} stabilized by any of the following symmetries: $\mathds{Z}_2\times\mathds{Z}_2$ rotation symmetry of the spins,  spatial inversion, and time-reversal symmetry. In the presence of $\mathds{Z}_2\times\mathds{Z}_2$, the phase is characterized by a nonlocal string order SO parameter~\cite{Pollmann12}, 
\begin{equation}
	\text{SO}(m,n) = \langle\psi_0| S_m^z e^{i\pi\sum_{p=m+1}^{n-1}S_p^z}S_n^z|\psi_0\rangle,
	\label{eq:SO}
\end{equation}
which approaches a finite, nonzero value in the thermodynamic limit $\text{SO} = \lim_{|n-m|\rightarrow\infty}\text{SO}(m,n)$. We also make use of another  non-local order parameter, $\mathcal{O}_{\mathcal{I}}$, which is based on the symmetry under spatial inversion and has been introduced in Ref. \onlinecite{Pollmann-2012b}. This order parameter  is basically a topological invariant which tells us to which cohomology-class the ground state belongs. It is directly obtainable from an iMPS in canonical form. The order parameter $\mathcal{O}_{\mathcal{I}}$ takes on values $+1$/$-1$ in the EH/OH phases respectively, and thus gives a clear distinction between these two phases. If the inversion symmetry is broken, the order parameter is set to $0$.

The field theory describing the XY phase is that of a free boson $\phi$. The XY order parameter, which has a ``stiffness'' that determines the energetic cost of modulating  $\phi$, is  parameterized here by the Luttinger parameter $K$.  Within the XY phase $K\geq 0.5$, leading to continuously varying exponents for the correlation functions, including one with a power-law decay
\begin{equation}
	\langle S_m^+ S_n^-\rangle \sim |n-m|^{-\alpha(\Delta,D_2,D_4)},
	\label{eq:SpmCorr}
\end{equation}
with correlation length $\xi_1$ and $S^\pm_n = S^x_n \pm iS^y_n$. The exponent $\alpha(\Delta,D_2,D_4)$ varies slowly over the XY phase. The $\langle S_m^z S_n^z\rangle$ correlation function is exponentially decaying, with correlation length $\xi_0$.\cite{Aschauer} In some parts of the phase diagram, the $S^z$ correlation function becomes a power-law, while the $S^\pm$ is exponentially decaying. For the parameters we focus on, $\Delta>0, D_2 \geq 0$, it is always the $S^\pm$ correlation function~\eqref{eq:SpmCorr} which is relevant. The excitation spectrum is gapless and the central charge in the entire XY phase is $c=1$, implying that the entanglement entropy $S_E$ diverges throughout the XY phase.

\paragraph*{Phase transitions} The AFM $\leftrightarrow$ EH phase transition is second order, except at large $\Delta$ where it turns first order. It is characterized by a continuous vanishing of the magnetic order parameter of the AFM phase and a diverging correlation length and entanglement entropy connected by $c = 1/2$. The first order transition is very different from the other phase transitions considered here. It does not have a critical behavior, no divergence in $S_E$, $\xi$ or $F$, and the critical parameters $c$ and $K$ are zero. However, it is characterized by a discontinuous slope of the ground state energy, from the crossing of two energy levels, and a discontinuous jump in the magnetic order parameter. The AFM $\leftrightarrow$ OH phase transition behaves in a similar way with the same characteristics along with a vanishing string order parameter of the OH phase. 

The EH $\leftrightarrow$ OH transition is a Gaussian transition described by a conformal field theory with central charge $c = 1$. In addition, the string order parameter~\eqref{eq:SO} vanishes continuously and $\mathcal{O}_{\mathcal{I}}$ changes abruptly when entering the EH phase. 

The phase transitions discussed so far are between gapped phases and the observables obtained from the DMRG output normally give a distinct signature at the transition. Transitions into the gapless XY phase are generally numerically more demanding.  Neither the EH nor the XY phase has an order parameter with a nonzero value in the thermodynamic limit, but transitions from the OH phase can in principle be determined from the vanishing of the string order parameter. Instead, one needs to rely on the characteristics of the transition, which is of Berezinskii-Kosterlitz-Thouless (BKT) type.~\cite{Berezinskii,Kosterlitz-Thouless} As the stiffness weakens, at a critical value of $K = 0.5$ vortices in the $\phi$ field unbind, the system becomes disordered and $K$ renormalizes to zero. At the BKT transition $K = 0.5$, however, logarithmic corrections to scaling are expected due to the presence of a marginal operator, which makes the transition difficult to study with finite size (or, with iDMRG, finite $\chi$) techniques. By measuring the stiffness $K$ and the central charge $c$, the BKT point can be determined by a sharp drop in $K$ at the value $K = 0.5$ and a sharp drop in $c$ from 1 (XY) to 0 (EH). 

\subsection{Results}
The results presented below were all obtained with the iDMRG algorithm described in Sec.~\ref{sec:MPS}.
We used a wide variety of bond dimensions $50\leq \chi\leq 1600$, see each specific data set below.
The effective correlation length obtained at criticality is $\xi_\chi \lesssim 6000$.

\subsubsection{\texorpdfstring{AFM $\leftrightarrow$ EH}{AFM, EH}}
\begin{figure}[tb!]
  \begin{center}
        \subfigure{\includegraphics[width=42mm]{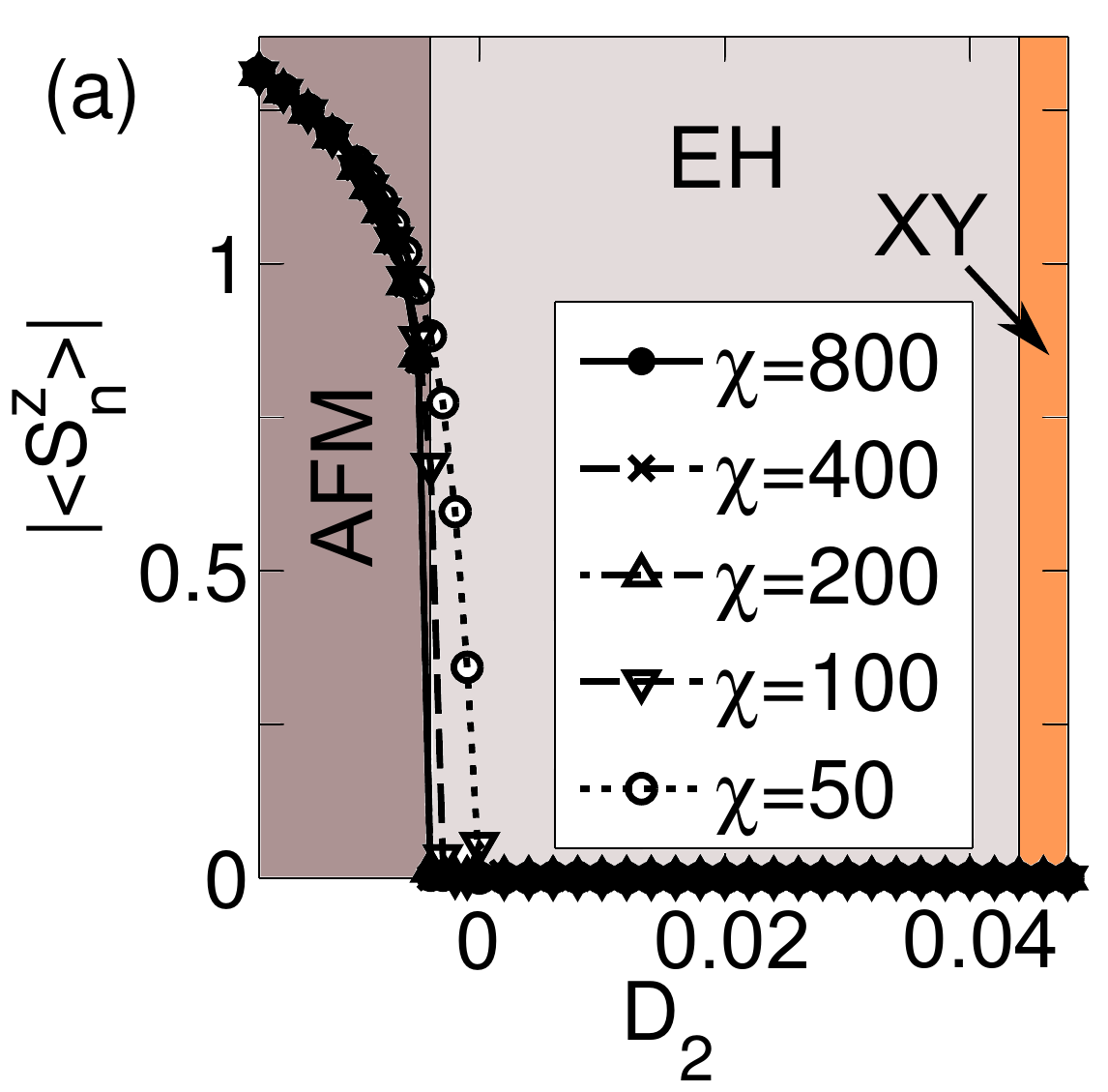}}    
            \subfigure{\includegraphics[width=42mm]{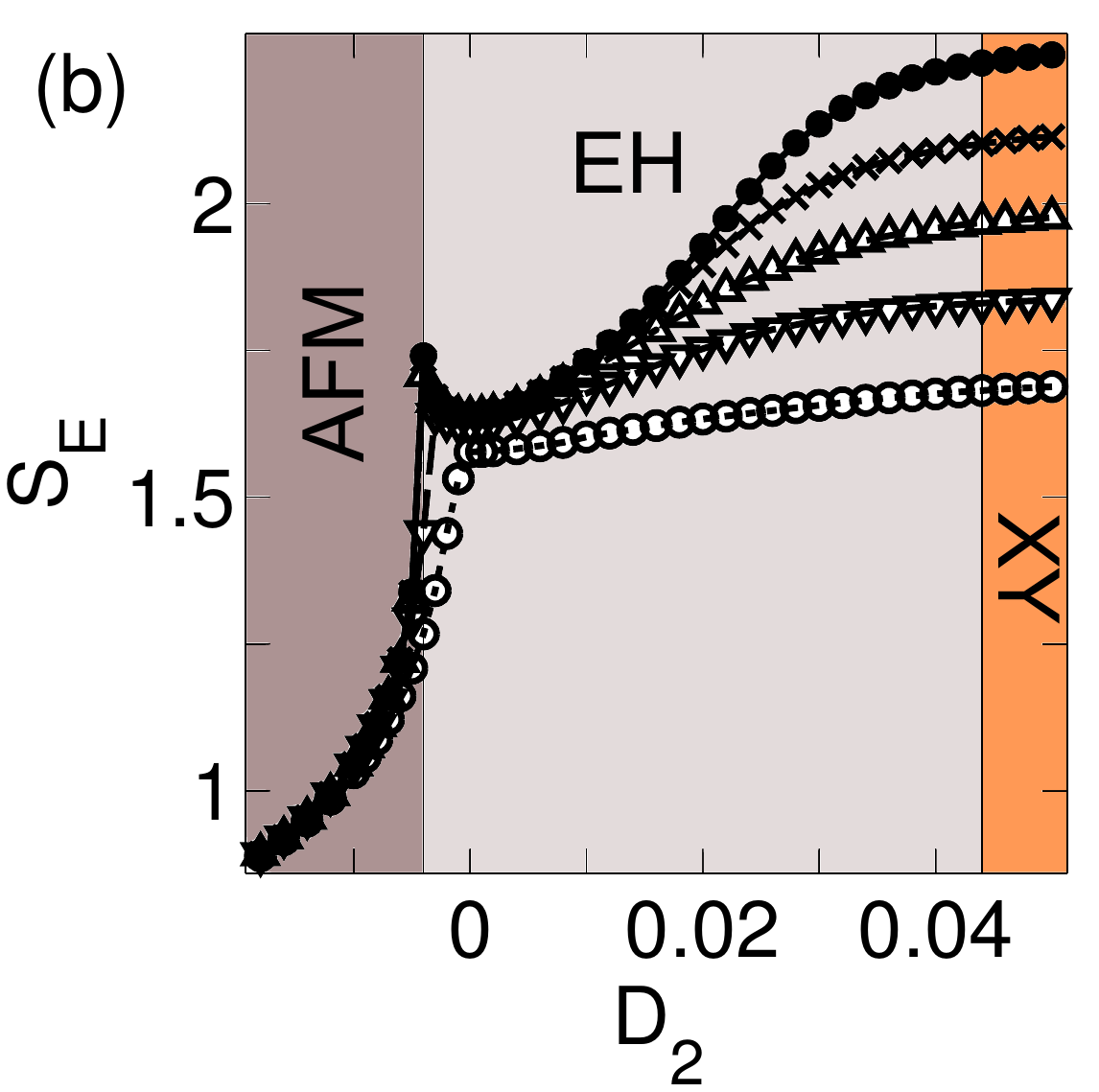}}\\
                \subfigure{\includegraphics[width=42mm]{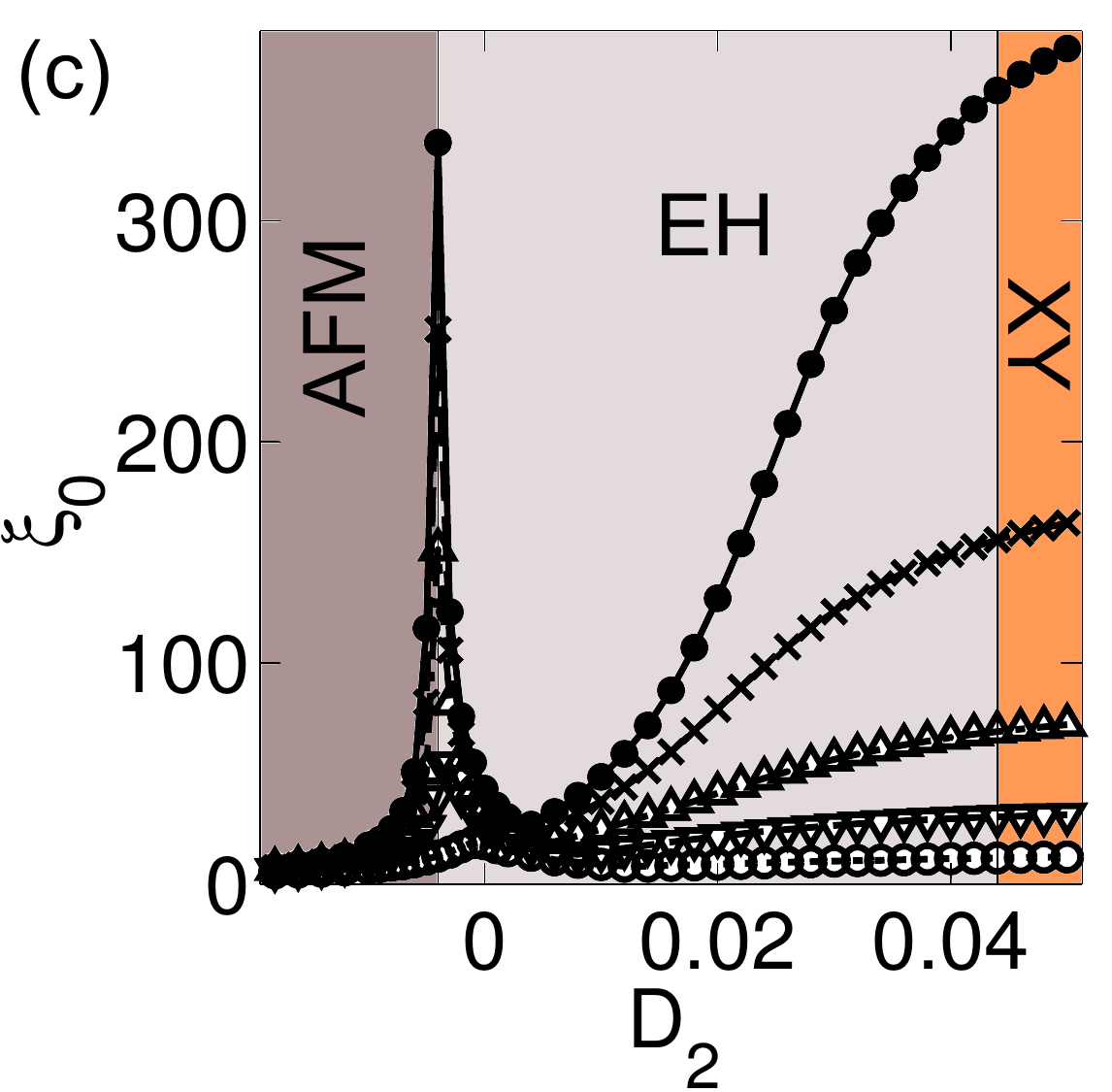}}
                       \subfigure{\includegraphics[width=42mm]{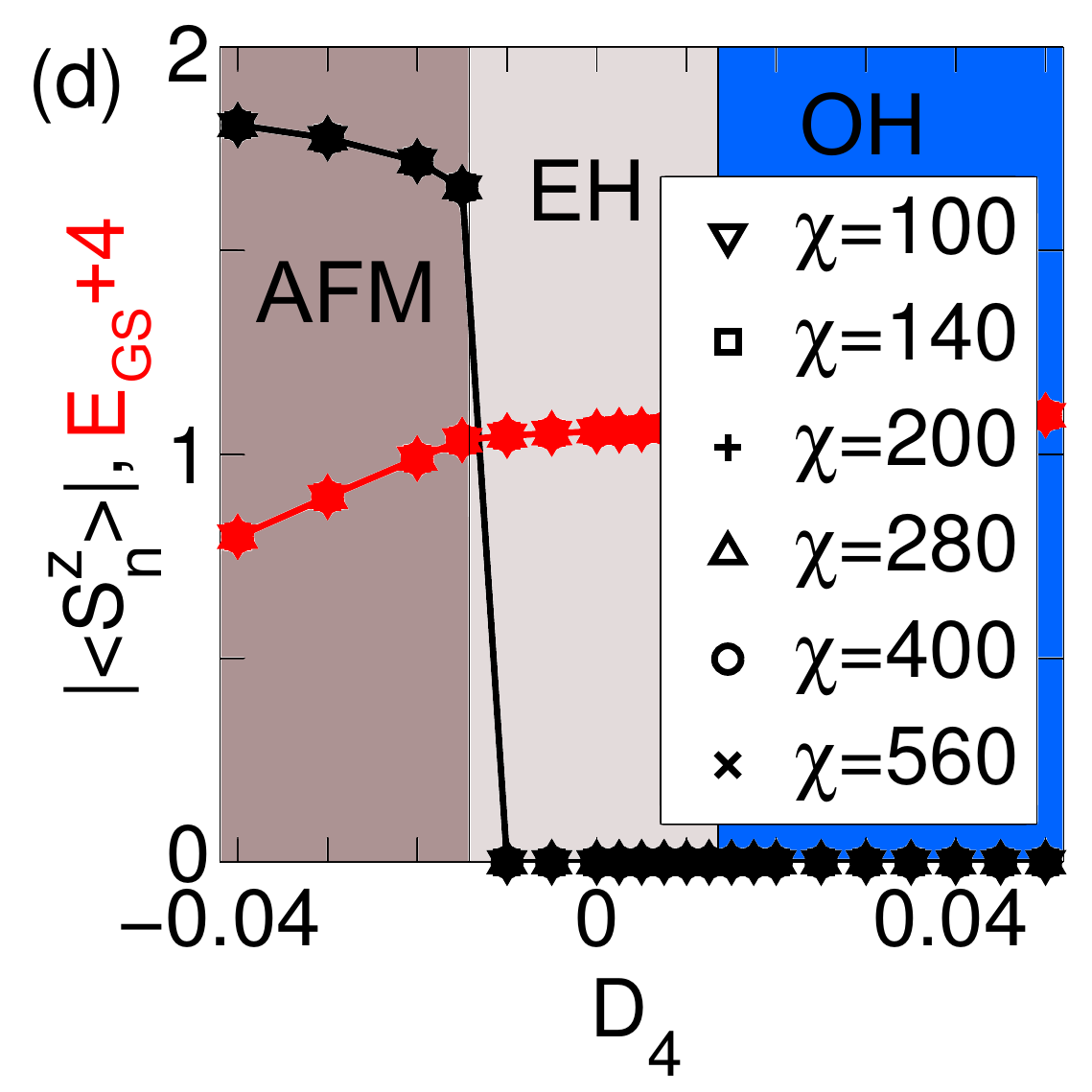}}
    \caption{(Color online) AFM $\leftrightarrow$ EH phase transitions. (a)-(c) Example of a 2$^\text{nd}$ order phase transition. Data as a function of $D_2$ away from the Heisenberg point, see Fig.~\ref{JzD2pd}. (a) The magnetic order parameter $|\langle S_n^z\rangle |$ of the AFM phase. (b) The entanglement entropy $S_E$. (c) The $\xi_0$ correlation length. (d) Example of a 1$^\text{st}$  order phase transition. Data along the line $D_2=3.95-D_4$ at $\Delta=4.5$, see Fig.~\ref{Gaus}(a). The magnetic order parameter $|\langle S_n^z\rangle |$ (black symbols) and the ground state energy $E_\text{GS}+4$ (red symbols) are plotted.}
    \label{AFMEH}
  \end{center}
\end{figure}

An example of data used to obtain the AFM~$\leftrightarrow$~EH phase boundary is presented in Fig.~\ref{AFMEH}. The local magnetization $|\braket{S_n^z}|$ is obtained from Eq.~(\ref{corrfcn}), the entanglement entropy $S_E$ from Eq.~(\ref{eq:SE}), the correlation length $\xi$ from Eq.~(\ref{eq:corr}), and the ground state energy $E_\text{GS}$ is defined in Sec.~\ref{sec:MPS}.D. The location of the 2$^\text{nd}$ order phase transition in (a)-(c) shifts with increasing $\chi$ and approaches the value $D_2^{\text{AFM-EH}}(\Delta=1,D_4=0)=-0.0046\pm0.0003$ at large $\chi$. However, both this shift and the scaling region (as defined in Fig.~\ref{fig:fes}) are much smaller than the narrow width of the EH phase around the Heisenberg point. This phase can hence be detected even with small $\chi$. Note though, that the shift of the phase transition location is large enough to give, at certain values of $D_2$, an increasing $|\langle S_n^z\rangle|$ with decreasing $\chi$, instead of the decreasing trend normally expected in the scaling region.

At $\Delta \approx 3.8$ the transition turns first order. In this case, the transition location is obtained from the kink in the ground state energy $E_\text{GS}$ and the discontinues vanishing of the magnetic order parameter, see Fig.~\ref{AFMEH}(d).  All data is well converged in $\chi$ in this type of phase transition without critical behavior. 

One of the main questions of this paper, whether the Heisenberg point is adiabatically connected to the large-$D$ region, relies on an accurate determination of the narrow part of the EH phase. In addition to locating the phase boundaries in Fig.~\ref{JzD2pd}, it is important to show that the Heisenberg point and the large-$D$ region do not belong to two different phases. To rule out a phase transition occurring between them, we note that while the scaling region surrounding a phase transition can be narrow in parameter space, a noticeable increase in quantities like $S_E$ and $\xi_{\text{phys}}$ is normally seen far away from a transition. Repeating simulations as in Fig.~\ref{AFMEH}(a)-(c) across the EH phase along lines spaced $\delta \Delta=0.1$ apart, we have found no such signatures. In particular, the minimum of $S_E$ and $\xi_{\text{phys}}$ remained roughly constant and at the same distance away from the AFM phase, all the way up to the change in phase transition type at $\Delta\approx 3.8$. Furthermore, no kink in $E_\text{GS}$ was observed. With no signs of a phase transition, an adiabatic connection between the Heisenberg point and the large-$D$ region is likely.

\subsubsection{\texorpdfstring{EH $\leftrightarrow$ XY}{EH, XY}}
\begin{figure}[tb!]
  \begin{center}
        \subfigure{\includegraphics[width=42mm]{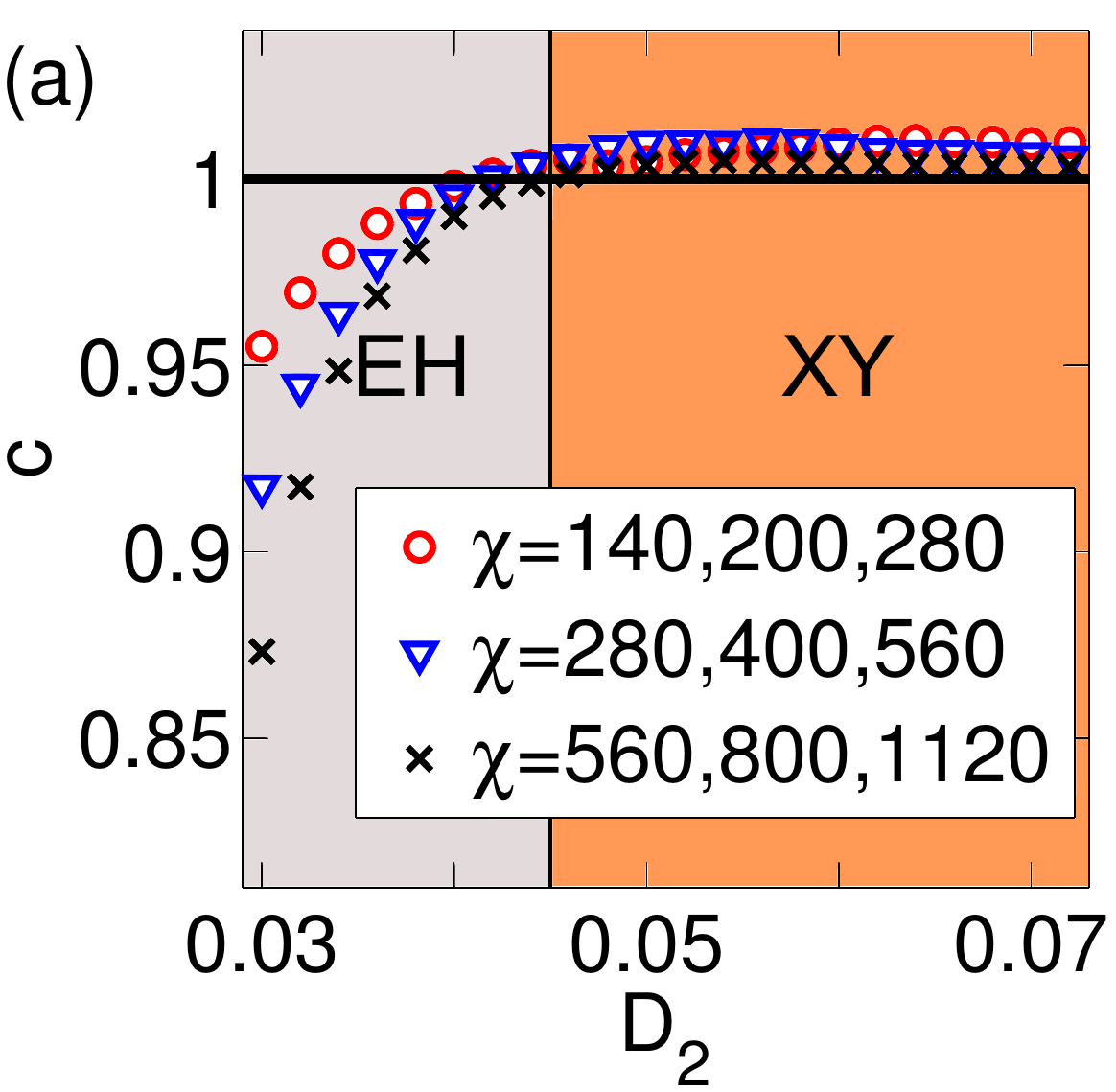}}    
            \subfigure{\includegraphics[width=42mm]{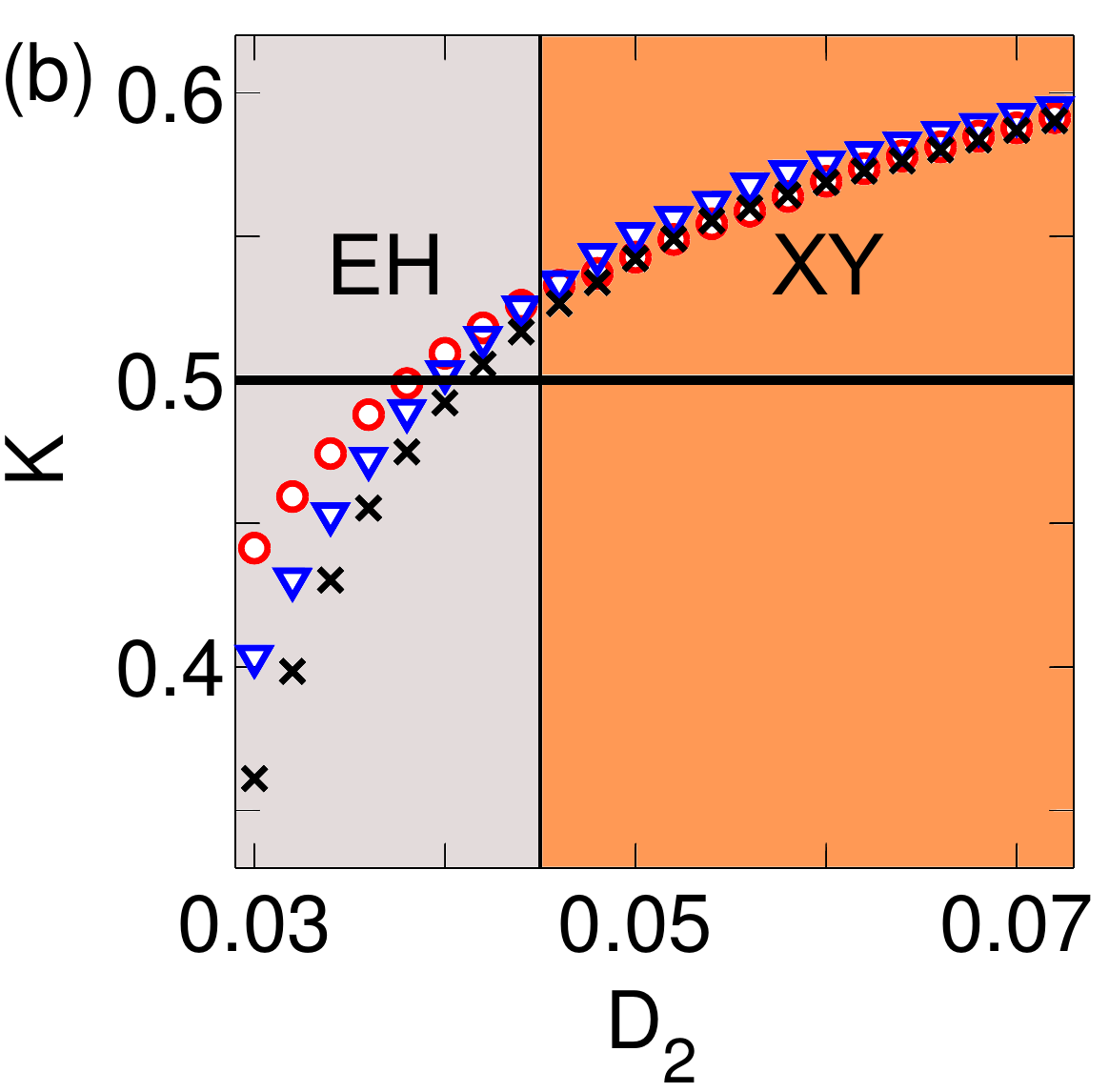}}\\
                \subfigure{\includegraphics[width=42mm]{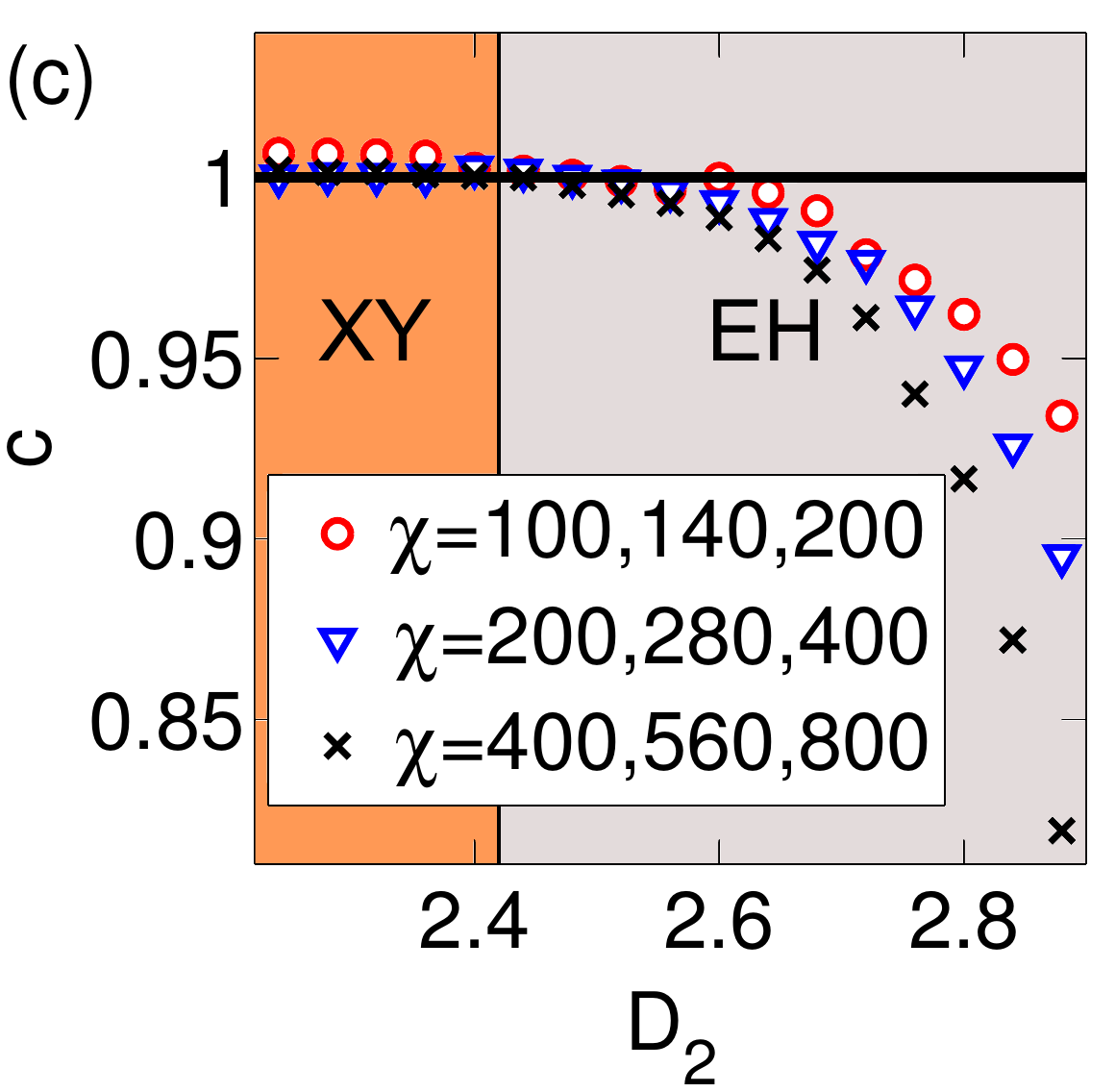}}
                       \subfigure{\includegraphics[width=42mm]{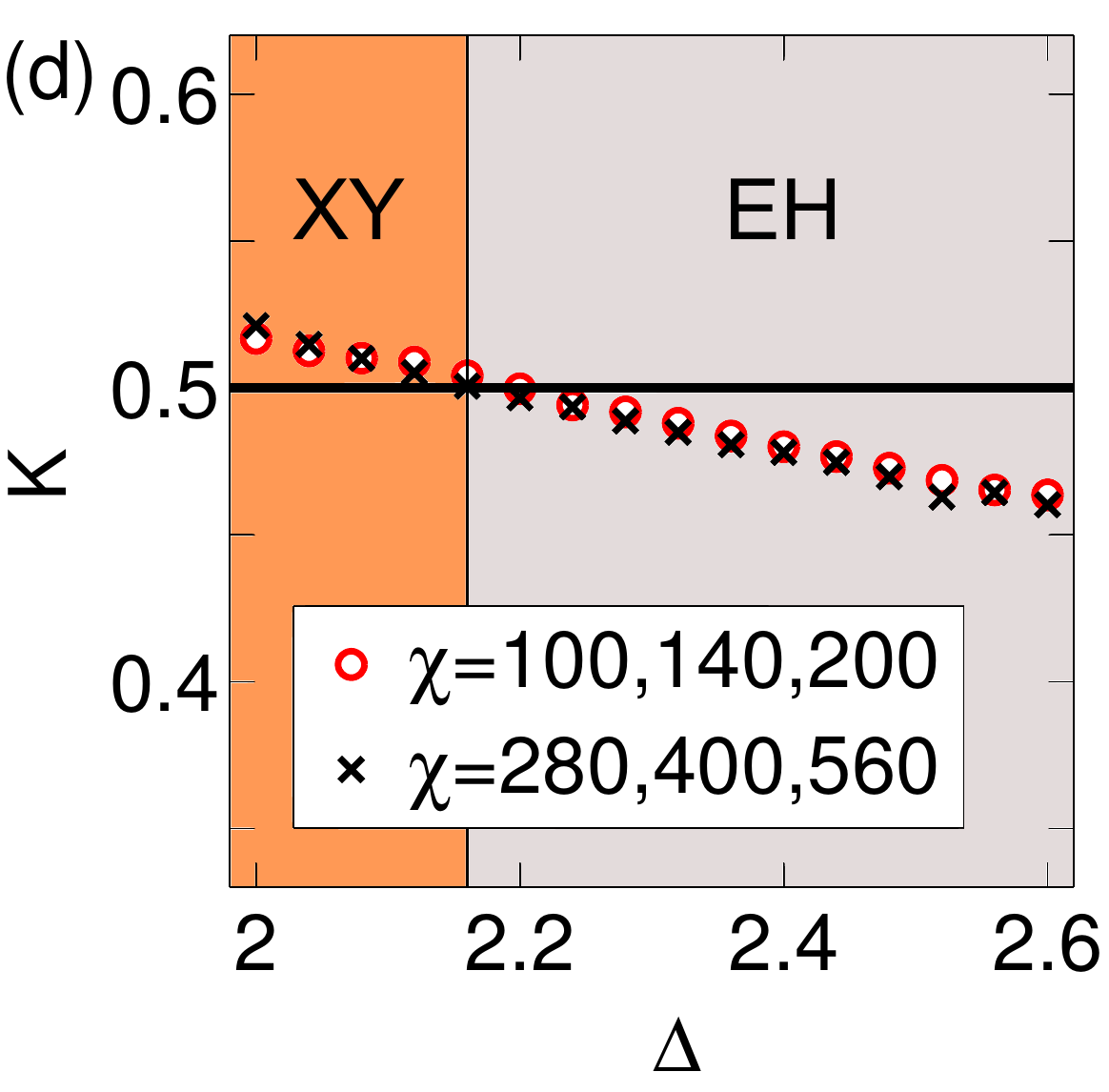}}         
    \caption{(Color online) BKT phase transition. The data are obtained from finite entanglement scaling averaged over the $\chi$ values given in the insets [for (b) the same $\chi$ values as in (a) were used]. (a)-(c) Critical quantities as a function of $D_2$ away from the Heisenberg point, see Fig.~\ref{JzD2pd}. (a) and (c) show the central charge, while (b) and (d) the stiffness. (d) Stiffness along the line $D_2=0.85\cdot\Delta-0.055$ at $D_4=0$, see Fig.~\ref{JzD2pd}.}
    \label{BKT}
  \end{center}
\end{figure}

Examples of the BKT type EH $\leftrightarrow$ XY phase transition, located with finite entanglement scaling, are shown in Fig.~\ref{BKT}. The central charge $c$ and the stiffness $K$ are obtained from a linear fit of the data at various values of $\chi$ to Eqs.~\eqref{tc} and~\eqref{Kc} respectively. Examples of the $S_E$ data can be found in Fig.~\ref{AFMEH}(b) and the data for $\xi=\xi_1\gg\xi_0$ in the XY phase and in the scaling region surrounding it behaves in a similar way to $\xi_0$ in Fig.~\ref{AFMEH}(c) in these regions. The nonzero values of $c$ and $K$ in the scaling region fall off continuously from their critical values to zero, both with increasing $\chi$ and with deviation of the coupling constant from criticality. The decrease is smallest near criticality and becomes more pronounced further away. However, small deviations in $c$  and $K$ from their critical values can be observed from data simulated with strict convergence criteria. 

Ideally, the phase transition is located by noting that in the EH phase $c$ and $K$ scale to zero with increasing $\chi$, while they scale to their critical values in the XY phase. This observation is complicated by the fact that just inside the XY phase the critical parameters, $c$ and especially $K$, have larger values than expected. This deviation decreases with increasing $\chi$ and disappears further into the critical phase. A similar observation was made by Song et al.~\cite{Song11} for $K$ at a BKT transition in a finite system, and was attributed to the presence of non-universal marginal operators at the phase transition causing finite size corrections to $K$. Similar marginal operators could give rise to finite entanglement correction in our case, however, in our data at the BKT transition these are small, due to the large $\chi$ used in the simulation. Nevertheless, due to these corrections and the relatively large variation in $K$ inside the XY phase, calculation of the central charge allows for a more accurate determination of the phase boundary than the stiffness calculations.

The BKT phase transition is more accurately determined with scaling in two observable that both depend on $\chi$, compared to scaling of a single observable with $\chi$. Examples of the former include $c$ or $K$ versus $\xi$ as we use here. Example of the latter are $S_E$ or $\xi_0$, [see data in Fig.~\ref{AFMEH}(b)-(c)], and energy gap $\delta E\propto 1/\xi$ as used in earlier DMRG studies~\cite{Schollwoeck95,Schollwoeck96,Aschauer} where a finite size scaling in $L$ was also required. Another method recently used to obtain the phase boundary in finite systems is level-spectroscopy (LS), which utilizes the crossing of excited energy levels in different magnetization sectors, with and without twisted boundary conditions.~\cite{Kitazawa,Nomura} In this method, an accurate determination of the BKT phase transitions is possible, even using finite systems with exact diagonalization (ED).~\cite{Nomura,Tonegawa,Tzeng} A comparison between our results for the phase transition location in Figs.~\ref{BKT}(a)-(b) and those of prior studies is given in Table~\ref{tab:BKT}. Good agreement is observed between the three different studies.

\begin{table}[tb!] 
\begin{center}
  \begin{tabular}{l c c}\hline\hline
       Method & \hspace{10mm}$D_2^\text{EH-XY} $ & \hspace{3mm}$D_2^\text{XY-EH}$ \\ \hline
       2OS+iDMRG & \hspace{10mm}$0.045\pm0.002$ & \hspace{3mm}$2.42\pm0.05$\\  
       LS+ED & \hspace{10mm}$0.043$ & \hspace{3mm}$2.39$\\
       1OS+DMRG & \hspace{10mm}$0.04\pm0.02$ & \hspace{3mm}$3.0\pm0.1$  \\  \hline \hline
         \end{tabular}
  \caption{Location of the BKT phase transition as a function of $D_2$ away from the Heisenberg point, calculated with three different methods. Our results are obtained from two observable scaling (2OS) and iDMRG, LS+ED results from ref.~\onlinecite{Nomura} and one observable scaling (1OS) and DMRG results from Refs.~\onlinecite{Schollwoeck95,Schollwoeck96}}   
  \label{tab:BKT}
\end{center}
\end{table}

Outside the XY phase, on the large $D_2$ side, see Fig.~\ref{JzD2pd}, the scaling region extends further than at small $D_2$, as can be seen in Fig.~\ref{BKT}(c) (note the different scale on the $D_2$ axis). A good agreement is again obtained between our results and the LS+ED study, see Table~\ref{tab:BKT}.  The agreement with the earlier DMRG studies is, however, not as good. The large scaling region, makes scaling in one observable, like $\delta E$ versus $\chi$, more unreliable than scaling in two observables and this accounts for most of the discrepancy.

Away from the Heisenberg point with increasing $\Delta$, the critical XY phase shrinks in the $(D_2,D_4)$ planes (see for example Fig.~\ref{D2D4pd}). $K\geq 0.5$ is only found for $\Delta\lesssim 2.2$, regardless of the on-site anisotropy strength. In the $D_4=0$ plane, stiffness calculations have $K=0.5$ extending to $\Delta \approx 2.18$. This agrees well with the LS+ED study of Ref.~\onlinecite{Tonegawa}. An example of the decrease in $K$ when leaving the XY phase along the line $D_2=0.85\Delta - 0.055$ at $D_4=0$ can be seen in Fig.~\ref{BKT}(d). Apart from the slow decrease of $K$ across the $K=0.5$ line, no other clear features can be observed, maybe with the exception of a finite $\chi$ scaling around $K=0.5$. However, this scaling is smaller than at the previous two BKT transitions we considered in Fig.~\ref{BKT} (data for $K$ not shown for the second of those), indicating that the XY phase could be slightly smaller in simulations with larger $\chi$ in this region. No decrease in $K$ with $\chi$ can be observed outside the XY phase. The reason is that our data can not distinguish these states from critical states with $c=1$, due to the nearby (in $D_4$) EH $\leftrightarrow$ OH phase transition. We discuss this in more detail below. Earlier DMRG results from the vanishing of the energy gap had a much larger XY phase, extending around this line, to $\Delta\approx 3.8$~\cite{Aschauer}. The close presence of the EH-OH phase transition is also the reason for the much larger XY phase that was obtained in earlier DMRG calculations.

\subsubsection{\texorpdfstring{EH $\leftrightarrow$ OH}{EH, OH}}
The second main question for the spin-2 chain we address, is whether the OH phase exists in the model~\eqref{Hamiltonian}. We argued in Sec.~\ref{sec:model} that with the addition of the $D_4$ term one effectively projects the system to a spin-1 system, realizing the corresponding spin-1 phase diagram. Indeed, as shown in Fig.~\ref{D2D4pd}, with increasing $D_4$ an OH phase with ground states that can be described accurately with an iMPS is observed in a large part of the phase diagram. The ground states have doubly degenerate entanglement spectra, non-zero string order, and $\mathcal{O}_\mathcal{I}=-1$. Below, we focus on exploring how close to the XXZ-chain the OH phase extends. Hence, we will primarily investigate the EH $\leftrightarrow$ OH phase transition for small on-site anisotropies, which is the experimentally most relevant case.

\begin{figure}[tb!]
  \begin{center}
            \includegraphics[width=42mm]{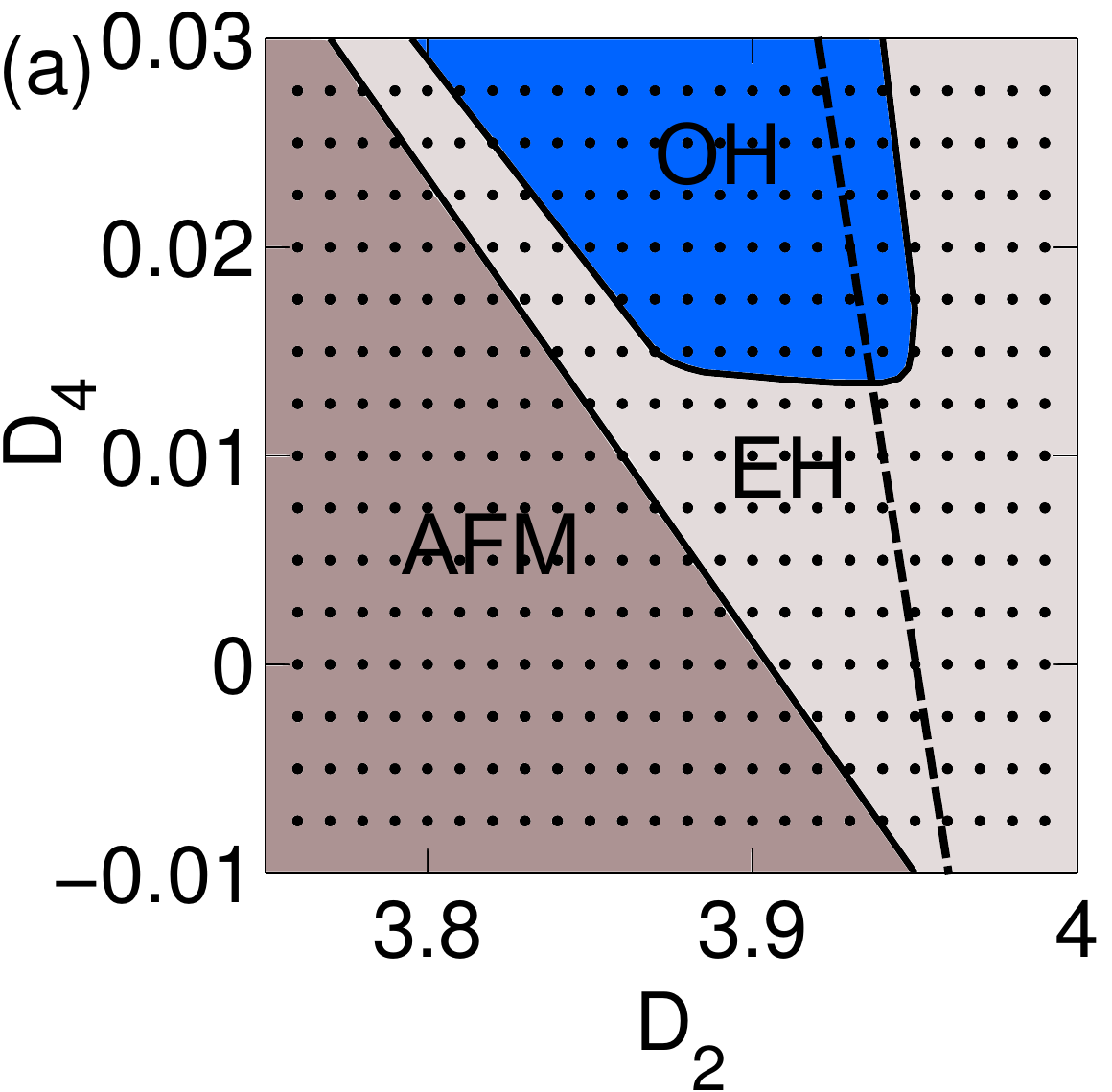}
              \includegraphics[width=42mm]{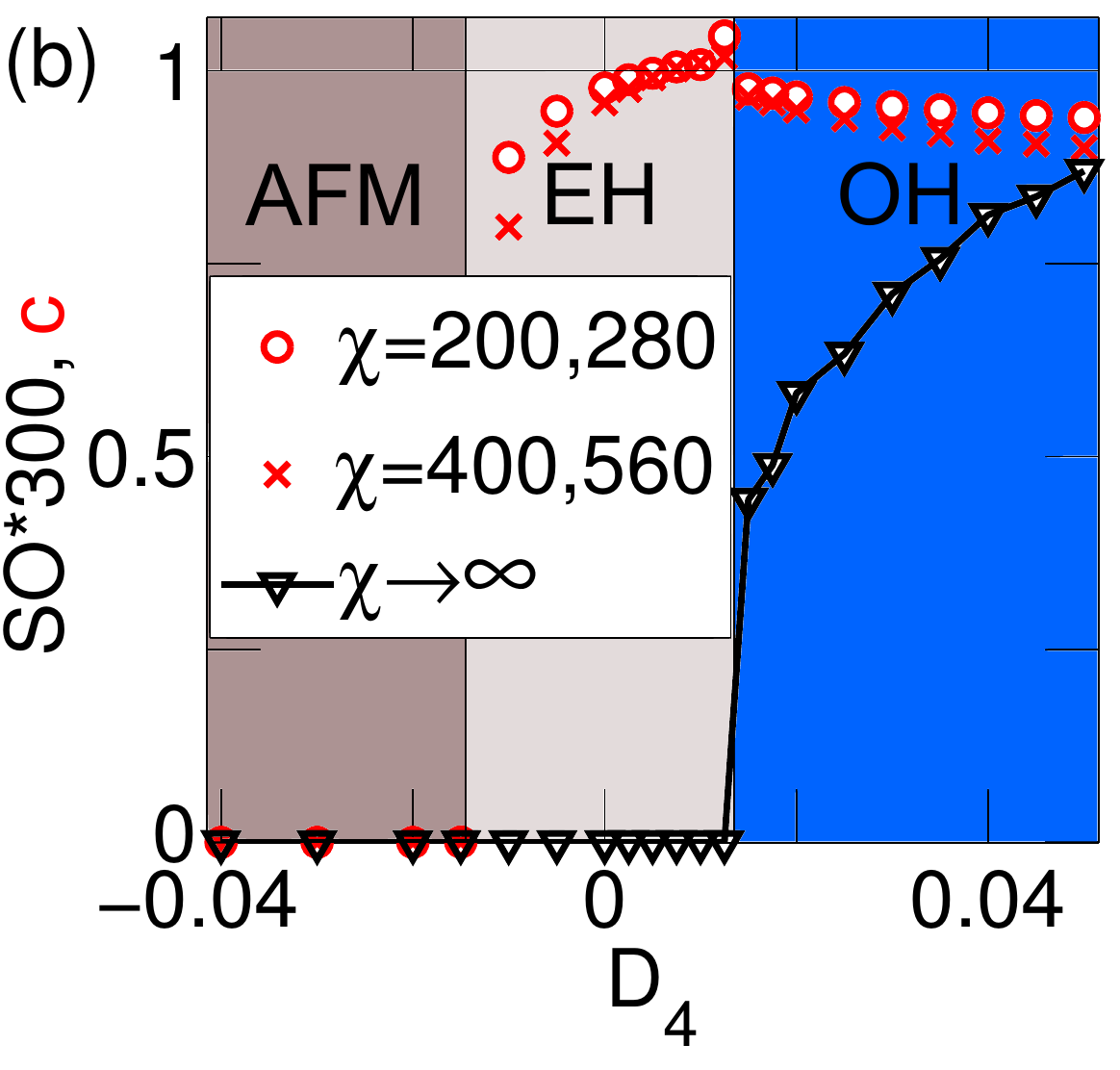}\\
               \includegraphics[width=42mm]{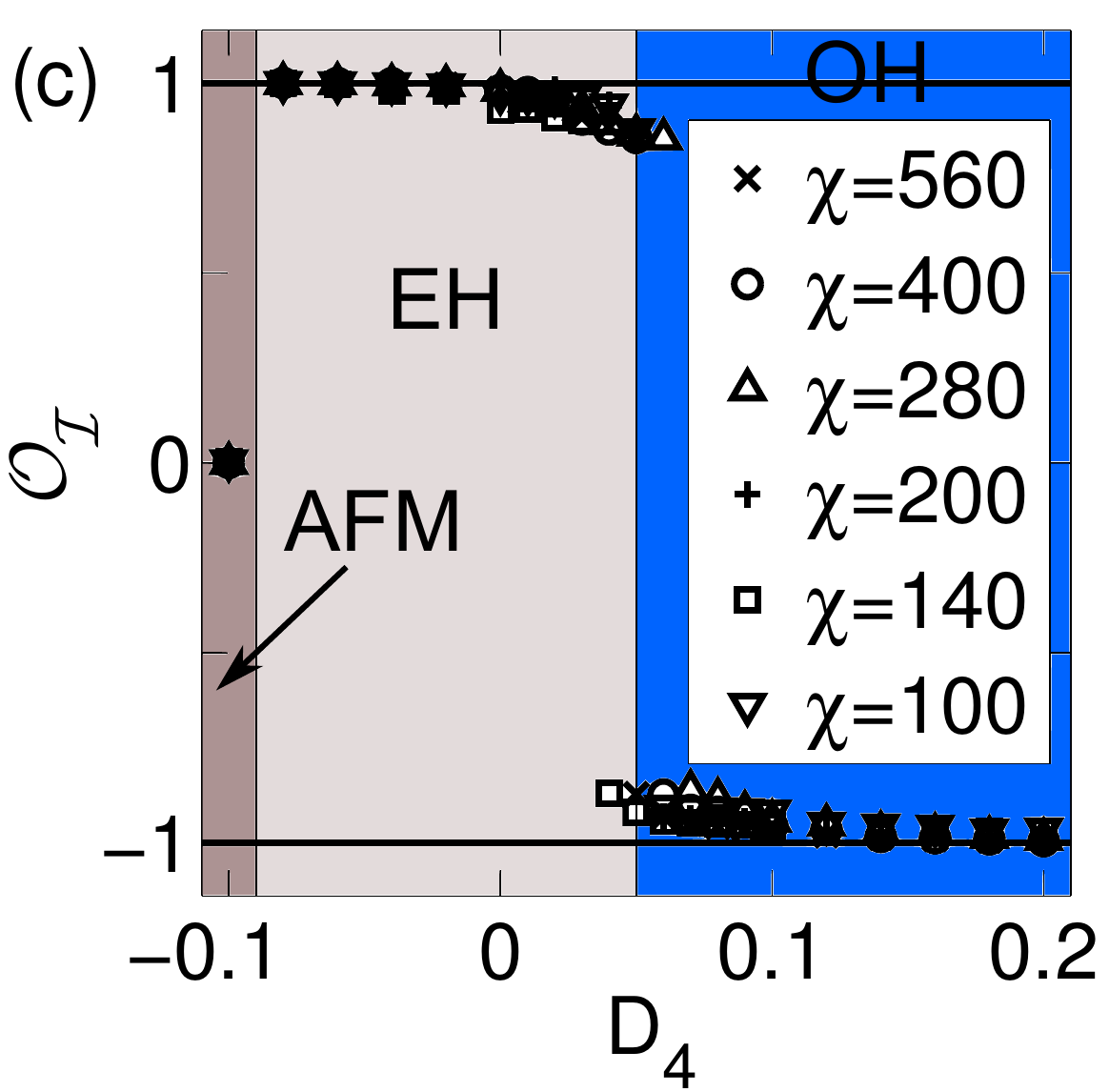}
                 \includegraphics[width=42mm]{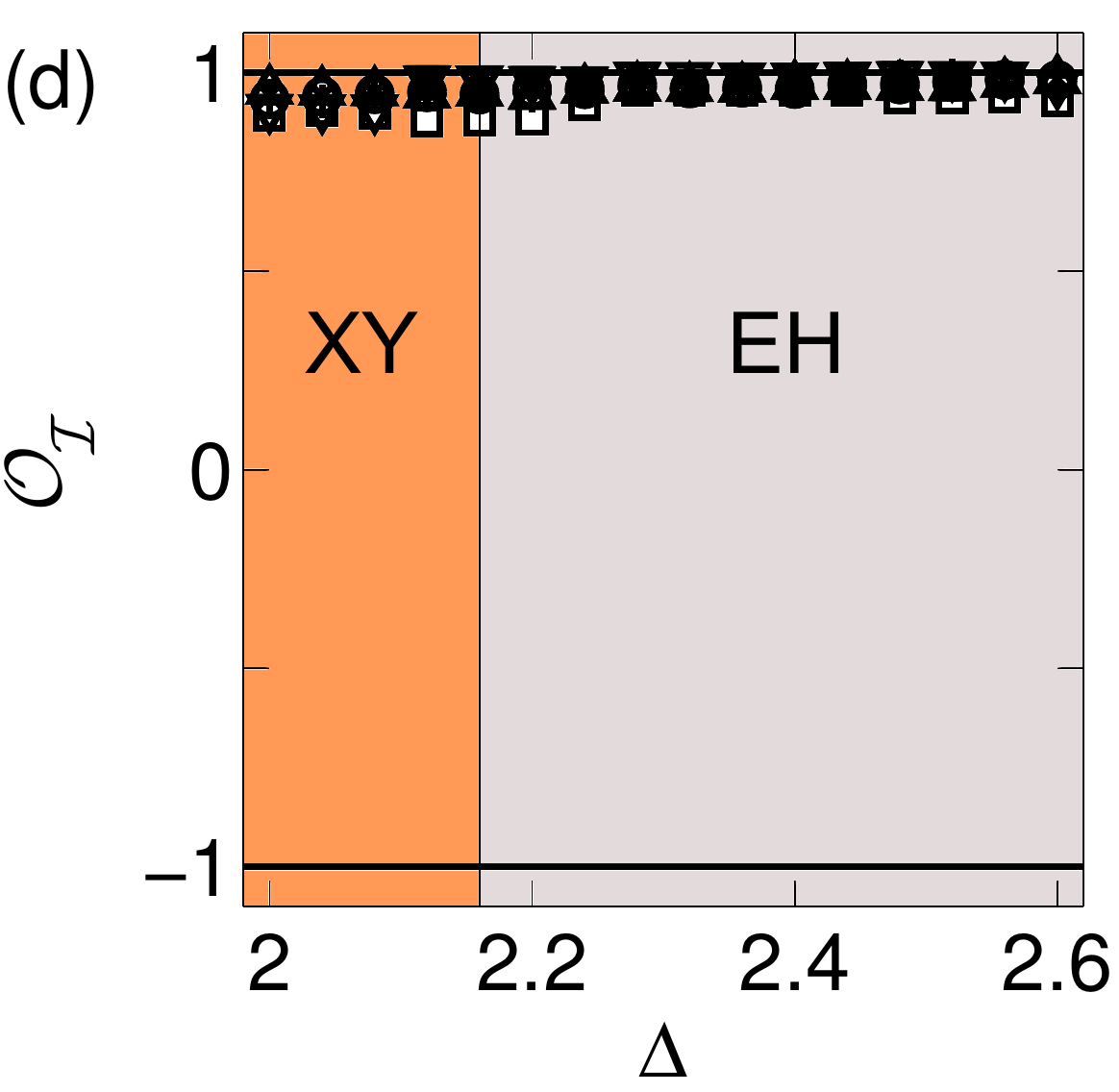} 
    \caption{(Color online) (a) Phase diagram at $\Delta=4.5$. (b) String order SO and central charge $c$ across the EH-OH phase transition along the line $D_2=3.95-D_4$ shown in (a). (c) Projected inversion symmetry along the line $D_2=2.155-D_4$ at $\Delta=2.6$. (d) Projected inversion symmetry $\mathcal{O}_\mathcal{I}$ along the line $D_2=0.85\cdot\Delta-0.055$ at $D_4=0$, see Fig.~\ref{JzD2pd}. 
		}
    \label{Gaus}
  \end{center}
\end{figure}

The OH phase extends to $\Delta^{\text{OH-EH}}\approx 5.3$ close to the AFM phase, much further than the XY phase which only extends to $\Delta\approx 2.2$. For example, the relevant part of the $(D_2,D_4)$ phase plane at $\Delta=4.5$ is shown in Fig.~\ref{Gaus}(a). The OH phase extends almost all the way to the $D_4=0$ plane. In fact, the transition into the EH phase at $D_4^{\text{EH-OH}}=0.0135\pm0.0010$ for $\Delta=4.5$ and $D_2=3.95$ is approximately the closest approach of the OH phase to the $D_4=0$ plane for any parameters in Eq.~\eqref{Hamiltonian}. 

The location of the EH $\leftrightarrow$ OH phase transition can be accurately determined from many different observables far away from the critical phase. Fig.~\ref{Gaus}(b) shows, for example, how the string order [obtained from Eqs.~\eqref{corrfcn} and~\eqref{nonlocalT} in the limit $\chi\rightarrow\infty$]  and the central charge vary across the phase transition along the line $D_2=3.95-D_4$ [see black dashed line in Fig.~\ref{Gaus}(a)]. The string order has a small value in the OH phase, and abruptly changes to zero at the phase transition. A peak in $c$ to a value slightly larger than 1, is observed at the phase transition consistent with the expected critical behavior. Since in this data in the OH phase, we are never very far from the phase transition, the decrease of $c$ with $\chi$ is slow.  The OH phase disappears again from this phase diagram [Fig.~\ref{Gaus}(a)] at large $D_4$ in the $D_4\rightarrow\infty$ limit. 

Closer to the critical XY phase, at smaller values of $\Delta$, it is harder to accurately locate the EH $\leftrightarrow$ OH phase transition. The peaks in the critical properties that diverge in the thermodynamic limit get broader and broader, for the $\chi$ we can simulate, while moving to a slightly larger $D_4$ as $\Delta$ is decreased. For $\Delta\lesssim 2.6$ it is unclear where the peak is and close to the XY phase $c\approx 1$ for $-0.03\lesssim D_4\lesssim 0.12$. The order parameters of the OH phase (SO and projected inversion symmetry) locate the phase transition more accurately. Fig.~\ref{Gaus}(c) shows $\mathcal{O}_\mathcal{I}$, calculated as in Ref.~\onlinecite{Pollmann-2012b}, along the line $D_2=2.155-D_4$ at $\Delta=2.6$. The phase transition is obtained at $D_4^{\text{EH-OH}}(\Delta=2.6,D_2=2.105)=0.05\pm0.04$. No scaling with $\chi$ is observed for this order parameter. 

Even closer to the XY phase both the expectation value and the uncertainty of the location of the EH $\leftrightarrow$ OH phase transition increase, the former to larger $D_4$, from similar data sets as in Fig.~\ref{Gaus}(c) (not shown). However, the uncertainty is always smaller than the distance to the $D_4=0$ plane as can be seen in Fig.~\ref{Gaus}(d), where no sign of an OH phase can be seen. The data is along the line $D_2=0.85\cdot\Delta-0.055$ at $D_4=0$ which is in the center of where Refs.~\onlinecite{Tonegawa,Tzeng} found a narrow OH phase. Also, the stiffness in Fig.~\ref{BKT}(d) was calculated along this line. 

Note, that no sign of the XY $\leftrightarrow$ EH phase transition can be seen in the $\mathcal{O}_\mathcal{I}$ order parameter. As the critical state cannot be represented exactly with a finite dimensional MPS, the order parameter   $\mathcal{O}_\mathcal{I}$ is not well defined here. The DMRG optimization  of the finite dimensional MPS, however, yields a state which has an approximate inversion symmetric ground state which lies in the EH phase with $\mathcal{O}_\mathcal{I}=1 $. 

Parts of the phase diagram of the spin-2 XXZ-chain with on-site anisotropy are difficult to study numerically. This is mainly due to the presence of the critical XY phase, the EH $\leftrightarrow$ OH phase transition which has some of the same critical properties ($c=1$), and the large scaling region that surrounds them. This is especially true around the line where these two phase transition types meet.  Earlier DMRG studies~\cite{Schollwoeck95,Schollwoeck96,Aschauer} could not determine the location of the phase boundaries accurately where the scaling region is large. While increased computational power helps, scaling in two observables that both depend on $\chi$ rather than a direct scaling with $\chi$ is even more important. In all other parts of the phase diagram our results are in good agreement with the above mentioned studies. 

The location of the BKT transition agrees well with that obtained in the LS+ED studies~\cite{Nomura,Tonegawa}. Some discrepancy between our result and LS the results~\cite{Tonegawa,Tzeng} can be found in the location of the EH $\leftrightarrow$ OH phase transition close to the XY phase, but the difference in the obtained values of $D_4$ is nevertheless small. The ED~\cite{Tonegawa} study considered small systems ($L\leq12$), and obtained a roughly linear scaling in $1/L^2$, just as what was obtained for the BKT phase transition. However, for larger systems ($L\leq 28$)~\cite{Tzeng} a different scaling was observed in DMRG leading to a smaller OH phase. In the region with three nearby competing phases and a huge correlation length $\xi_\text{phys}\gtrsim 10000$, scaling from small systems can sometimes be unreliable. Even some of our data is uncertain in this region, since a bond dimension of $\chi\leq560$ ($800$ for some points) is not enough to clearly distinguish between the many nearly degenerate energy eigenstates present. 

Finally, we mention that a recent study\cite{TuOrus} on a related model showed that the OH phase also appears in a spin-2 chain at the SO(5) symmetric point, obtained by tuning the $J_p$'s in $H=\sum_n\sum_{p=1}^4 J_p(\vec{S}_n \cdot \vec{S}_{n+1})^p$. 

\section{spin-3 XXZ-chain}
\label{sec:S3}
Having applied the iDMRG algorithm to study the phase diagram of the spin-2 XXZ Heisenberg model, we provide the first, to our knowledge, numerical investigation of the spin-3 XXZ-chain~\eqref{XXZ} around the Heisenberg point. The approach to the classical limit $S\rightarrow\infty$ is analyzed by comparing the results with the $S=1$ and $2$ cases. The spin-3 chain is hard to investigate numerically, not primarily because of the large local Hilbert space, but due to the tiny OH phase with the correspondingly large correlation lengths, see Sec.~\ref{sec:model}. 

In the semi-classical limit of large but finite $S$, an exponentially decreasing gap is obtained at the Heisenberg point: $\delta E_S \sim S^2e^{-\pi S}$ as $S\rightarrow \infty$, with a spin wave velocity $v=\delta E_S \xi=2S$ relating it to the correlation length.~\cite{HaldaneA,HaldaneB} The width of the Haldane phase in $\Delta$ also decrease rapidly with $S$ in the semi-classical limit.

\begin{figure}[tb!]
  \begin{center}
    \subfigure{\includegraphics[width=42mm]{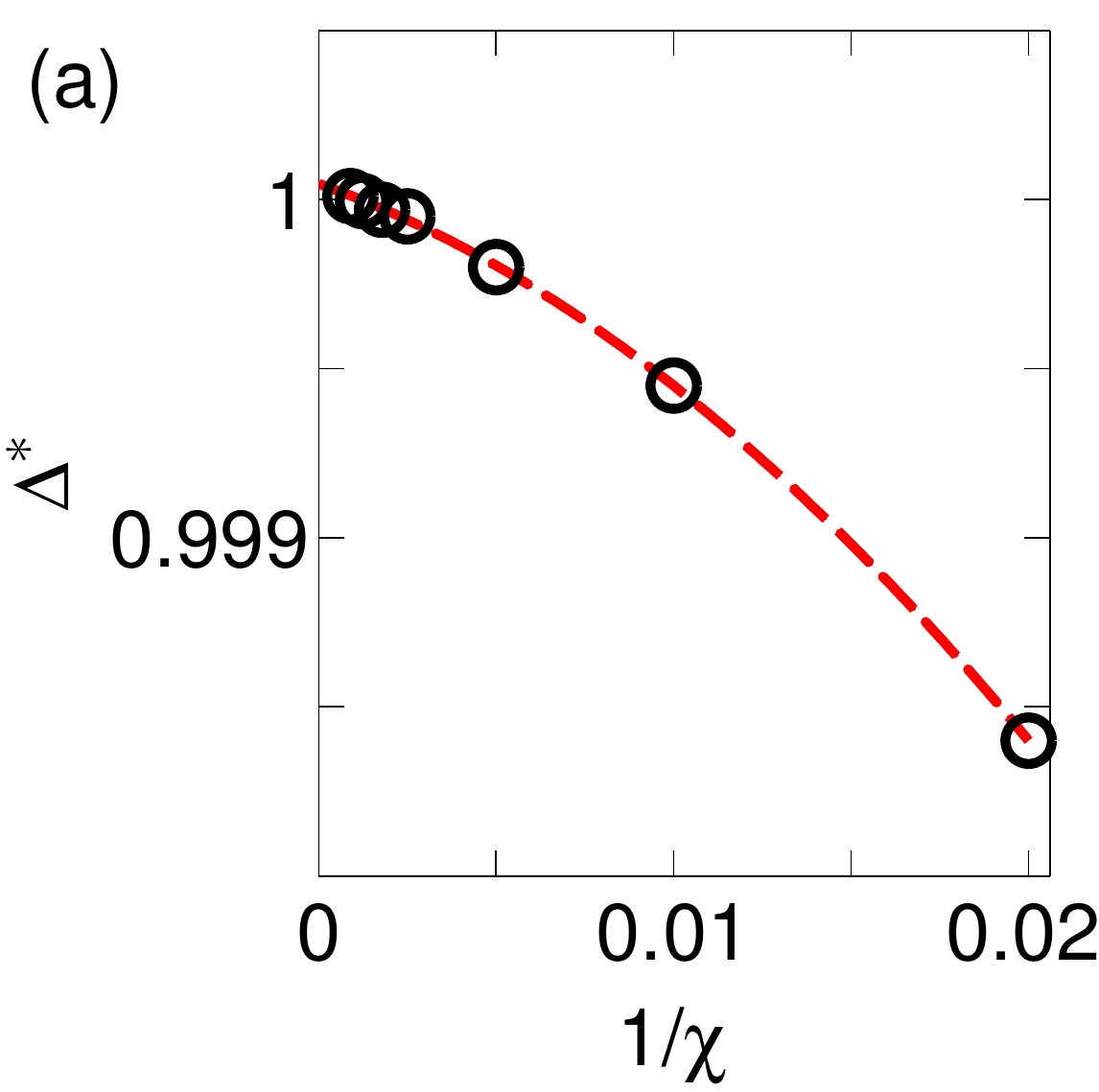}}
        \subfigure{\includegraphics[width=42mm]{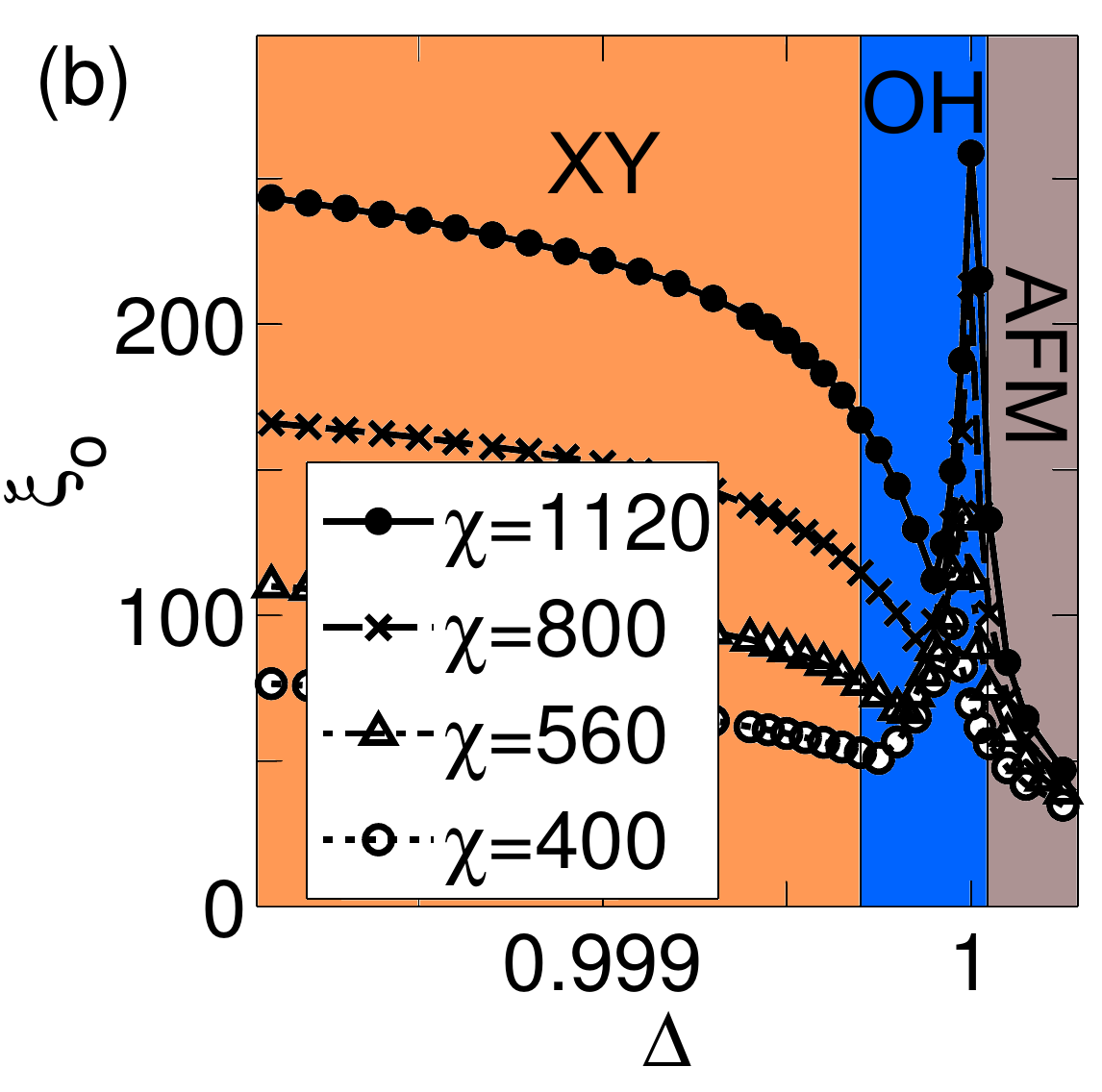}}\\    
            \subfigure{\includegraphics[width=42mm]{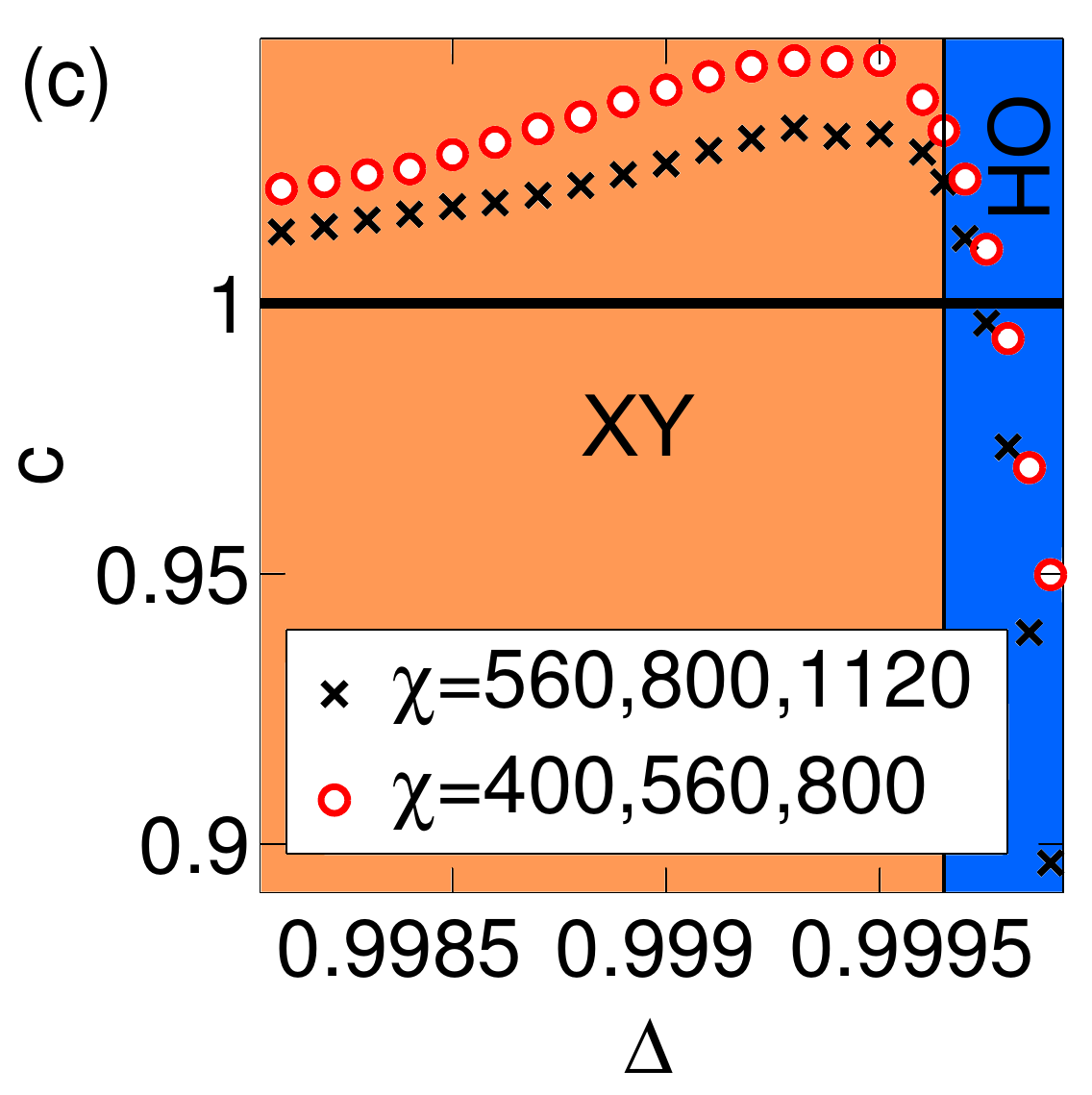}}
                \subfigure{\includegraphics[width=42mm]{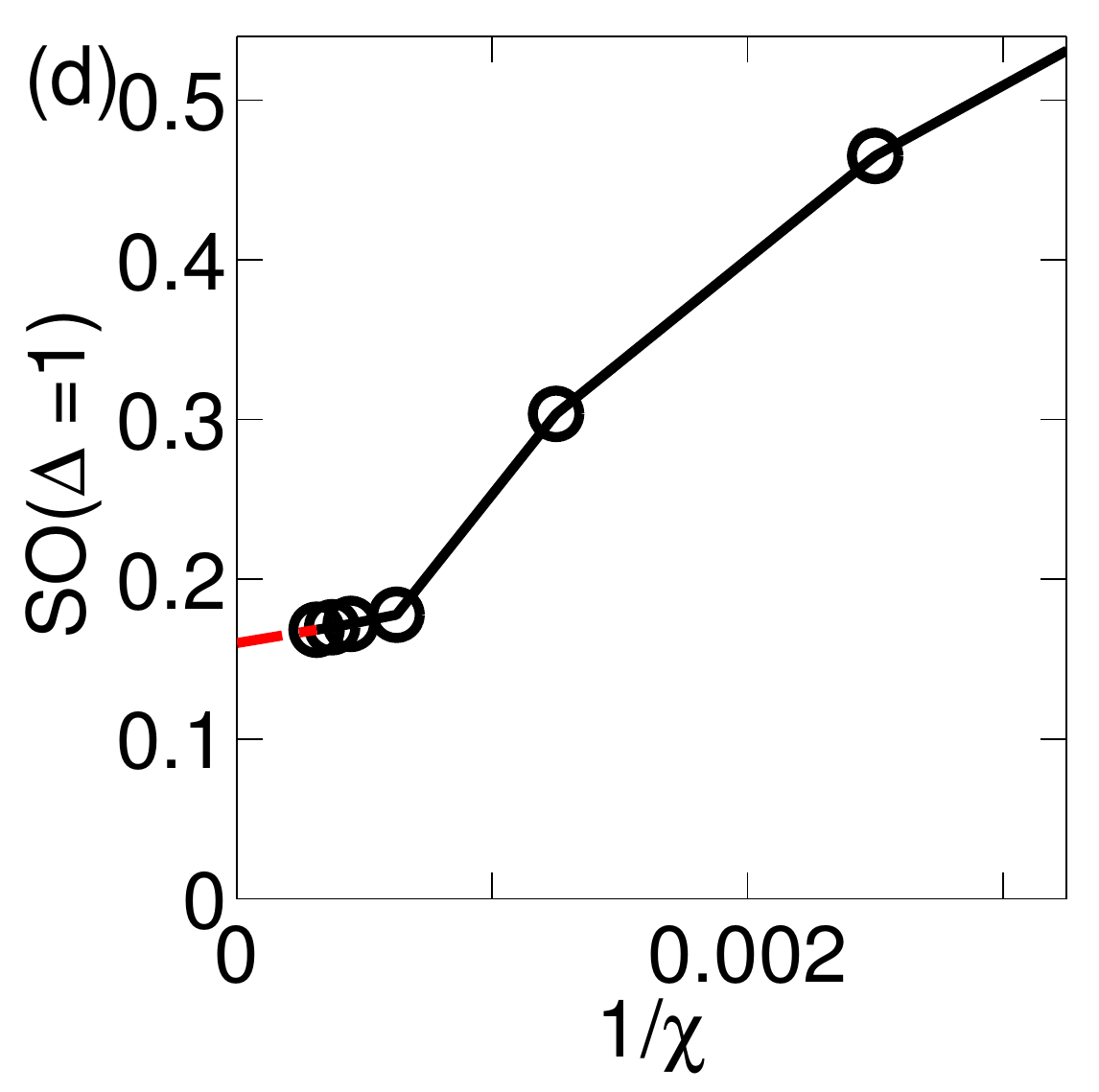}}
    \caption{(Color online) Spin-3 XXZ-chain. (a)  The magnetic order parameter vanishes $|\langle S^z_n\rangle |\rightarrow 0$ at different $\Delta^*$ as a function of the $\chi$ used in the simulation. (b) The correlation length $\xi_0$ for various $\chi$ across the XY $\leftrightarrow$ OH and the OH $\leftrightarrow$ AFM phase transitions. (c) The central charge, fitted to data at three different $\chi$, across the  XY $\leftrightarrow$ OH phase transition. (d) String order at the Heisenberg point ($\Delta=1.0$) for various $\chi$ and scaled to the thermodynamic limit (red dashed line).}
    \label{S3}
  \end{center}
\end{figure}

For $S=1,2$ the size of the AFM phase is slightly overestimated at finite $\chi$ but decreases with increasing $\chi$, see Fig.~\ref{AFMEH}(a). The same occurs for $S=3$, see Fig.~\ref{S3}(a), which plots the value $\Delta^*$ of $\Delta$ at which the AFM magnetic order parameter vanishes ($\braket{S_n^z}\rightarrow 0$) as a function of $1/\chi$, for $\chi\leq 1120$. In the thermodynamic limit, the OH $\leftrightarrow$ AFM phase transition is at $\Delta_{S=3}^{\text{OH-AFM}}=1.000045\pm0.000020$, very close to the Heisenberg point. The shift of the phase boundary with increasing $\chi$ is also observed in the peak of the $\xi_0$ correlation length, see Fig.~\ref{S3}(b). 

The XY $\leftrightarrow$ OH phase transition at $\Delta_{S=3}^{\text{XY-OH}}=0.99965\pm0.00010$ is located by the vanishing of the string order (not shown) and central charge calculations, see Fig.~\ref{S3}(c). Locating the point at which the string order vanishes is challenging. The string order decays with a power-law in the critical phase and its value in the thermodynamic limit in the gapped OH phase approaches zero exponentially at the XY $\leftrightarrow$ OH phase transition. However, with a careful scaling of the string order versus $\chi$ the same phase transition location as with the central charge is obtained. The projected inversion symmetry $\mathcal{O}_\mathcal{I}$ is not useful here since it is undefined at criticality and can not detect this type of phase transition, see Fig.~\ref{Gaus}(d) and its discussion. 

At the Heisenberg point, the OH phase is obtained as the ground state for $\chi\gtrsim 800$, as can be seen from the point at which the curve in Fig.~\ref{S3}(a) crosses $\Delta^* = 1$. The correlation length $\xi^{S=3}=520\pm60$ and the string order $\text{SO}^{S=3}=0.162\pm0.002$, see Fig.~\ref{S3}(d), in the thermodynamic limit is hence obtained from the data at $\chi=800,1120,1340,1600$ by scaling. Comparing with $S=1,2$ where $\xi^{S=1}=5.77\pm0.01$ and $\xi^{S=2}=47.2\pm0.5$, we indeed have an exponential growth, but with a bit smaller exponent than in the asymptotic limit $S\rightarrow\infty$. Nevertheless, it is large enough to make the Haldane phases in the quantum $S\geq4$ XXZ-chains unreachable with the current approach. Moreover, the width of the Haldane phase decrease even faster than $1/\xi$, with the  XY $\leftrightarrow$ Haldane phase transition $\Delta_{S}^{\text{XY-H}}$ and the Haldane $\leftrightarrow$ AFM phase transition $\Delta_{S}^{\text{H-AFM}}$ presented in Table~\ref{tab:S3} for $S=1,2,3$. For spin-3 the narrow OH phase is thus not observed for $\chi\lesssim 200$.

\begin{table}[tb!] 
\begin{center}
  \begin{tabular}{c c c}\hline\hline
       Spin & \hspace{10mm}$\Delta^\text{XY-OH} $ & \hspace{3mm}$\Delta^\text{OH-AFM}$ \\ \hline
       1 & \hspace{10mm}$0.0$ & \hspace{3mm}$1.19\pm0.01$\\  
       2 & \hspace{10mm}$0.962\pm0.004$ & \hspace{3mm}$1.0039\pm0.0005$\\
       3 & \hspace{10mm}$0.99965\pm0.00010$ & \hspace{3mm}$1.000045\pm0.000020$  \\  \hline \hline
         \end{tabular}
				 \caption{Location of the phase transitions in and out of the Haldane phase in $S=1,2,3$ XXZ-chains. For $S=1,2$ this data has been obtained before, for example in Ref.~\onlinecite{Kitazawa96}. They are given here within a wide interval for a comparison to the spin-3 results.} 
  \label{tab:S3}
\end{center}
\end{table}

\section{Conclusions}
In this work we have carefully studied the phase diagram of the spin-2 XXZ Heisenberg model with on-site anisotropies using the iDMRG algorithm. We have established that the SPTP OH phase is obtained for not too large values of the on-site anisotropy, and that the Heisenberg point and the large-$D$ region are adiabatically connected and belong to the EH phase. We have also provided the first, to our knowledge, study of the corresponding spin-3 XXZ Heisenberg model. Thereby we have numerically further verified the exponentially decreasing gap and size of the Haldane phase with increasing spin. Many of these conclusions were greatly aided by scaling in two observables that both vary with the bond dimension $\chi$, which can locate deviations from criticality more accurately than scaling directly with $\chi$. Additionally, we have provided a self-contained and didactic introduction to the iDMRG algorithm, which was applied in this study.

\section*{Acknowledgments}
We acknowledge Joel E. Moore, Luis Seabra, and Stephan Rachel for useful conversations, and especially Masaki Oshikawa for suggesting the inclusion of the $D_4$ term in our study. This work is supported by the ARO Optical Lattice Emulator program (J.K.), der Max-Planck-Gesellschaft (J.K. \& F.P.), NSF GRFP Grant DGE 1106400 (M.P.Z.), NSF DMR-0804413 and Sherman Fairchild Foundation (R.M.), and DOE BES DMSE (J.H.B.).


%

\end{document}